# Transparent Anonymization: Thwarting Adversaries Who Know the Algorithm

XIAOKUI XIAO

Nanyang Technological University

YUFEI TAO

Chinese University of Hong Kong

and

NICK KOUDAS

University of Toronto

---

Numerous generalization techniques have been proposed for privacy preserving data publishing. Most existing techniques, however, implicitly assume that the adversary knows little about the anonymization algorithm adopted by the data publisher. Consequently, they cannot guard against privacy attacks that exploit various characteristics of the anonymization mechanism. This paper provides a practical solution to the above problem. First, we propose an analytical model for evaluating disclosure risks, when an adversary knows *everything* in the anonymization process, except the sensitive values. Based on this model, we develop a privacy principle, *transparent l-diversity*, which ensures privacy protection against such powerful adversaries. We identify three algorithms that achieve transparent *l*-diversity, and verify their effectiveness and efficiency through extensive experiments with real data.

---

## 1. INTRODUCTION

Privacy protection is highly important in the publication of sensitive personal information (referred to as *microdata*), such as census data and medical records. A common practice in anonymization is to remove the identifiers (e.g., social security

---

This work was supported by the Nanyang Technological University under SUG Grant M58020016 and AcRF Tier 1 Grant RG 35/09, and by the Hong Kong Research Grants Council under GRF Grants 4169/09, 4173/08, and 4161/07.

Author's address: X. Xiao, School of Computer Engineering, Nanyang Technological University, Singapore 639798; email: xkxiao@ntu.edu.sg; Y. Tao, Department of Computer Science and Engineering, Chinese University of Hong Kong, Shatin, Hong Kong; email: taoyf@cse.cuhk.edu.hk; N. Koudas, Department of Computer Science, University of Toronto, Toronto, ON M5S2E4, Canada; email: koudas@cs.toronto.edu.

| Name | Age | Zipcode | Disease |
|------|-----|---------|---------|
| Ann | 21 | 10000 | dyspepsia |
| Bob | 27 | 18000 | flu |
| Cate | 32 | 35000 | gastritis |
| Don | 32 | 35000 | bronchitis |
| Ed | 54 | 60000 | gastritis |
| Fred | 60 | 63000 | flu |
| Gill | 60 | 63000 | dyspepsia |
| Hera | 60 | 63000 | diabetes |

Table I.    Microdata $T_1$

| Name | Age | Zipcode |
|------|-----|---------|
| Ann | 21 | 10000 |
| Bob | 27 | 18000 |
| *Bruce* | 29 | 19000 |
| Cate | 32 | 35000 |
| Don | 32 | 35000 |
| Ed | 54 | 60000 |
| Fred | 60 | 63000 |
| Gill | 60 | 63000 |
| Hera | 60 | 63000 |

Table II.    Voter List $E_1$

| Age | Zipcode | Disease |
|-----|---------|---------|
| [21, 27] | [10k, 18k] | dyspepsia |
| [21, 27] | [10k, 18k] | flu |
| 32 | 35000 | gastritis |
| 32 | 35000 | bronchitis |
| [54, 60] | [60k, 63k] | gastritis |
| [54, 60] | [60k, 63k] | flu |
| [54, 60] | [60k, 63k] | dyspepsia |
| [54, 60] | [60k, 63k] | diabetes |

Table III.    Generalization $T_2^*$

numbers or names) that uniquely determine entities of interest. This, however, is not sufficient because an adversary may utilize the remaining attributes to identify individuals [Samarati 2001]. For instance, consider that a hospital publishes the microdata in Table I, without disclosing the patient names. Utilizing the publicly-accessible voter registration list in Table II, an adversary can still discover Ann's disease, by joining Tables I and II. The joining attributes {*Age, Zipcode*} are called the *quasi-identifiers* (QI).

*Generalization* [Samarati 2001] is a popular solution to the above problem. It works by first assigning tuples to *QI-groups*, and then transforming the QI values in each group to an identical form. As an example, Table III illustrates a generalized version of Table I with three QI-groups. Specifically, the first, second, and third QI-groups contain the tuples {Ann, Bob}, {Cate, Don}, and {Ed, Fred, Gill, Hera}, respectively. Even with the voter registration list in Table II, an adversary still cannot decide whether Ann owns the first or second tuple in Table III, i.e., Ann's disease cannot be inferred with absolute certainty.

Generalizations can be divided into *global recoding* and *local recoding* [LeFevre et al. 2005]. The former demands that if two tuples have identical QI values, they must be generalized to the same QI-group. Without this constraint, the generalization is said to use local recoding. For instance, Table III obeys global recoding. Notice that Cate and Don have equivalent QI-values in the microdata (Table I), and therefore must be included in the same QI-group. This is also true for Fred, Gill, and Hera.

The privacy-preservation power of generalization relies on the underlying *privacy principle*, which determines what is a publishable QI-group. Numerous principles are available in the literature, offering different degrees of privacy protection. One popular, intuitive and effective principle is *l-diversity* [Machanavajjhala et al. 2007]. It requires that, in each QI-group, at most $1/l$ of the tuples can have the same sensitive value[1]. This ensures that an adversary can have at most $1/l$ confidence in inferring the sensitive information of an individual. For example, Table III is 2-diverse. Thus, an adversary can discover the disease of a person with at most 50% probability.

Interestingly, none of the existing privacy principles (except those in [Wong et al. 2007] and [Zhang et al. 2007]) specifies any requirement on the algorithm that produce the generalized tables. Instead, they impose constraints only on the formation

---

[1]There also exist other formulations of *l*-diversity, as will be discussed in Section 2.1

of the QI-groups (like $l$-diversity does), which, unfortunately, leaves open the opportunity for an adversary to breach privacy by exploiting the characteristics of the generalization algorithm. This problem is first pointed out by Wong et al. [2007], who demonstrate a *minimality attack*[2] that (i) can compromise most existing generalization techniques, and (ii) requires only a small amount of knowledge about the generalization algorithm. As a solution, they propose an anonymization approach that can guard against minimality attacks.

The work by Wong et al. reveals an essential issue in publishing microdata: a generalization method should preserve privacy, even against adversaries with knowledge of the anonymization algorithm. Towards addressing this issue, the techniques in [Wong et al. 2007] establish the first step by dealing with minimality attacks, which, however, is still insufficient for privacy protection. Specifically, given information about the anonymization method, an adversary can easily devise other types of attacks to circumvent a generalized table. To explain this, in the following we first clarify how minimality attacks work, and then, elaborate the deficiencies of [Wong et al. 2007].

**Minimality Attacks.** Good generalization should keep the QI values as accurate as possible. Towards this objective, the previous algorithms [Bayardo and Agrawal 2005; Fung et al. 2005; Ghinita et al. 2007; LeFevre et al. 2005; 2006a; Xiao and Tao 2007] produce *minimal generalizations*, where no QI-group can be divided into smaller groups without violating the underlying privacy principle. For example, Table III is a minimal 2-diverse generalization of Table I under global recoding. In particular, the first (second) QI-group in Table III cannot be divided, since any split of the group results in two QI-groups with a single tuple, which apparently cannot be 2-diverse. On the other hand, as Fred, Gill, and Hera have identical QI values, their tuples must be in the same QI-group, as demanded by global recoding. Therefore, the only way to partition the third QI-group is to break it into {Ed} and {Fred, Gill, Hera}, which also violate 2-diversity.

Minimal generalizations can lead to severe privacy breach. Consider that a hospital holds the microdata in Table IV, and releases the 2-diverse Table V, which is a minimal generalization under global recoding. Assume that an adversary has access to the voter registration list in Table II. Then, s/he can easily identify the six individuals in the second QI-group $G_2 = \{$Cate, Don, Ed, Fred, Gill, Hera$\}$ in Table V. After that, the adversary can infer the diseases of Cate and Don by reasoning as follows (i.e., a minimality attack). First, there exist only two tuples in $G_2$ with the same disease, which is *gastritis*. Second, since Table V is minimal, if we split $G_2$ into two parts $G_3 = \{$Cate, Don$\}$ and $G_4 = \{$Ed, Fred, Gill, Hera$\}$, either $G_3$ or $G_4$ must violate 2-diversity. Assume that $G_4$ is not 2-diverse. In that case, at least three tuples in $G_4$ should have an identical sensitive value, contradicting the fact that, in $G_2$, the maximum number of tuples with the same *Disease* value is 2. It follows that $G_3$ cannot be 2-diverse, indicating that both Cate and Don have the same disease, which must be *gastritis* (as mentioned earlier, no other disease is

---

[2]Note that minimality attack can be effective only when the microdata is anonymized with generalization or a similar methodology called *anatomy* [Xiao and Tao 2006a]. There exist other anonymization methods that are immune to attacks based on knowledge of the anonymization algorithm, as will be discussed in Section 5.

| Name | Age | Zipcode | Disease |
|------|-----|---------|---------|
| Ann | 21 | 10000 | dyspepsia |
| Bob | 27 | 18000 | flu |
| Cate | 32 | 35000 | gastritis |
| Don | 32 | 35000 | gastritis |
| Ed | 54 | 60000 | bronchitis |
| Fred | 60 | 63000 | flu |
| Gill | 60 | 63000 | dyspepsia |
| Hera | 60 | 63000 | diabetes |

Table IV.    Microdata $T_3$

| Age | Zipcode | Disease |
|-----|---------|---------|
| [21, 27] | [10k, 18k] | dyspepsia |
| [21, 27] | [10k, 18k] | flu |
| [32, 60] | [35k, 63k] | gastritis |
| [32, 60] | [35k, 63k] | gastritis |
| [32, 60] | [35k, 63k] | bronchitis |
| [32, 60] | [35k, 63k] | flu |
| [32, 60] | [35k, 63k] | dyspepsia |
| [32, 60] | [35k, 63k] | diabetes |

Table V.    Generalization $T_4^*$

**Algorithm** *Vul-Gen* $(T)$
1. if $T$ is the microdata $T_1$ in Table I
    return the generalization $T_4^*$ in Table V
2. otherwise, return a generalization of $T$ that is different from $T_4^*$

Fig. 1.    The *Vul-Gen* algorithm

possessed by two tuples in $G_2$).

**Motivation.** Wong et al. [2007] advance the other solutions by assuming that an adversary has one extra piece of knowledge: *whether* the anonymization algorithm produces a minimal generalization (note: the adversary is not allowed to have other details of the algorithm). Under this assumption, minimality attacks can be prevented using a simple solution — just deploy non-minimal generalizations. Nevertheless, given knowledge of the algorithm, can the adversary employ other types of attacks to compromise non-minimal generalizations? The answer, unfortunately, is positive, as can be demonstrated in a simple example as follows.

EXAMPLE 1. Consider the conceptual anonymization algorithm *Vul-Gen* in Figure 1. The algorithm takes as input a microdata table $T$, and generates a generalization $T^*$ of $T$. In particular, *Vul-Gen* outputs the generalization $T_4^*$ in Table V, if and only if $T$ equals the microdata $T_1$ in Table I. Notice that, $T_4^*$ is not a minimal 2-diverse version of $T_1$. This is because, the second QI-group of $T_4^*$, including the tuples {Cate, Don, Ed, Fred, Gill, Hera}, can be divided into 2-diverse QI-groups {Cate, Don} and {Ed, Fred, Gill, Hera}, which conform to global recoding.

Assume that a data publisher applies *Vul-Gen* on $T_1$, and releases the resulting 2-diverse generalization $T_4^*$. Since $T_4^*$ is not minimal, it does not suffer from minimality attacks. However, imagine an adversary who knows that *Vul-Gen* is the generalization algorithm adopted by the publisher. Once $T_4^*$ is released, the adversary immediately concludes that $T_1$ is the microdata, because *Vul-Gen* outputs $T_4^*$ if and only if the input is $T_1$. Hence, the adversary learns the exact disease of every individual, i.e., releasing $T_4^*$ causes a severe privacy breach.    □

It is clear from the above discussion that preventing minimality attacks alone is insufficient for privacy preservation, since an adversary (with understanding about the generalization algorithm) may employ numerous other types of attacks to infer sensitive information. This leads to a challenging problem: how can we anonymize the microdata in a way that proactively prevents *all* privacy attacks that may be launched based on *knowledge of the algorithm*?

Zhang et al. [2007] present the first theoretical study on the above problem. The core of their solution is a privacy model in which the anonymization algorithm (adopted by the publisher) is assumed to be public knowledge[3]. As will be discussed in Section 2.3, however, Zhang et al.'s privacy model is only applicable on a small subset of anonymization algorithms that (i) are deterministic, (ii) adopt global recoding generalization, and (iii) follow a particular algorithmic framework. This severely restricts the design of new anonymization approaches under the model, and makes it impossible to verify the privacy guarantees of existing randomized or local-recoding-based algorithms. Furthermore, the anonymization algorithms proposed by Zhang et al. have high time complexities: All but one algorithm run in time exponential in the number $n$ of tuples in the microdata, while the remaining one has a time complexity that is polynomial in $n$ and the total number $m$ of possible generalizations of the microdata. Note that, in practice, $m$ can be an exponential of $n$, since there may exist an exponential number of ways to divide the tuples in the microdata into QI-groups. As a consequence, the algorithms developed by Zhang et al. are rather inapplicable in practice.

**Contributions.** This paper develops a practical solution for data publishing against an adversary who knows the anonymization algorithm. First, we propose a model for evaluating the degree of privacy protection achieved by an anonymized table, assuming that the adversary has knowledge of (i) the anonymization algorithm employed by the publisher, (ii) the algorithmic parameters with which the anonymized table is computed, and (iii) the QI values of all individuals in the microdata. Our model captures all deterministic and randomized generalization algorithms [Aggarwal et al. 2006; Bayardo and Agrawal 2005; Fung et al. 2005; Ghinita et al. 2007; LeFevre et al. 2005; 2006a; 2006b; Iyengar 2002; Wang et al. 2004; Wong et al. 2006; Xiao and Tao 2007; Xu et al. 2006; Wong et al. 2007; Zhang et al. 2007], regardless of whether they adopt global recoding or local recoding. The model is even applicable for anonymized tables produced from *anatomy* [Xiao and Tao 2006a], a popular anonymization methodology that will be clarified in Section 2.1. Based on this model, we develop a new privacy principle called *transparent l-diversity*, which safeguards privacy against the adversary we consider.

As a second step, we identify two sufficient conditions for transparent $l$-diversity, based on which we propose three anonymization algorithms that achieve transparent $l$-diversity. None of these algorithms could have been possible under Zhang et al.'s privacy model, as they are either randomized or based on local recoding. We provide detailed analysis on the characteristics of each algorithm, and show that they all run in $O(n^2 \log n)$ time. In addition, we demonstrate the effectiveness and efficiency of our algorithms through extensive experiments with real data. Compared with the existing anonymization techniques that do not ensure transparent $l$-diversity, our solutions not only provide stronger privacy protection, but also achieve satisfactory performance in terms of data utility and computation overhead.

The rest of the paper is organized as follows. Section 2 presents the theoretical

---

framework that underlies transparent $l$-diversity. Section 3 presents our generalization algorithms, which are experimentally evaluated in Section 4. Section 5 surveys the previous work related to ours. Finally, Section 6 concludes the paper with directions for future research.

## 2. PRIVACY MODEL

This section presents our analytical model for assessing disclosure risks. In Section 2.1, we formalize several basic concepts. After that, Section 2.2 elaborates the derivation of disclosure risks. Section 2.3 discusses the differences between our model and the methods in [Wong et al. 2007] and [Zhang et al. 2007].

### 2.1 Preliminaries

Let $T$ be a microdata table to be published. We assume that $T$ contains $d + 2$ attributes, namely, (i) an identifier attribute $A^{id}$, which is the primary key of $T$, (ii) a sensitive attribute $A^s$, and (iii) $d$ QI attributes $A_1^q, ..., A_d^q$. As in most existing work, we require that $A^s$ should be categorical, while the other attributes can be either numerical or categorical.

For each tuple $t$ in $T$, let $t[A]$ be the value of $t$ on the attribute $A$. We define a *QI-group* as a set of tuples, and a *partition* of $T$ as a set of disjoint QI-groups of $T$ whose union equals $T$. We say that two QI-groups $G_1$ and $G_2$ are *isomorphic*, if (i) $G_1$ and $G_2$ contain the same multi-set of sensitive values, and (ii) every tuple $t_1 \in G_1$ shares the same identifier and QI values with a tuple $t_2 \in G_2$, and vice versa. For instance, let $G_1$ be a QI-group that contains the first two tuples in Table I. Suppose that we swap the sensitive values of Ann and Bob, such that Ann (Bob) has *flu* (*dyspepsia*). Then, the resulting QI-group $G_2$ is isomorphic to $G_1$.

We formalize the *anonymization* of $T$ as follows.

DEFINITION 1 (ANONYMIZATION). *An **anonymization function** $f$ is a function that maps a QI-group to another set of tuples, such that for any two isomorphic QI-groups $G_1$ and $G_2$, $f(G_1) = f(G_2)$ always holds. Given a partition $P$ of $T$ and an anonymization function $f$, a table $T^*$ is an **anonymization** of $T$ decided by $P$ and $f$, if and only if $T^* = \bigcup_{G \in P} f(G)$.*

There exist two popular types of anonymization methodologies, namely, generalization [Samarati 2001] and anatomy [Xiao and Tao 2006a]. Specifically, generalization employs an anonymization function that maps a QI-group $G$ to a set $G^*$ of tuples, such that (i) for any tuple $t^* \in G^*$, $t^*[A_i^q]$ ($i \in [1, d]$) is an interval containing all $A_i^q$ values in $G$, and (ii) any two tuples in $G^*$ have the same QI values. Anatomy, on the other hand, adopts an anonymization function that transforms a QI-group $G$ to two separate sets of tuples, such that first (second) set contains only the QI (sensitive) values in $G$. For example, given a partition of Table I that contains three QI-groups {Ann, Bob}, {Cate, Don}, and {Ed, Fred, Gill, Hera}, Table VI illustrates an anonymization of Table I produced from anatomy. Observe that Table VIa (VIb) contains only the QI (sensitive) values in Table I.

The techniques developed in this paper can be incorporated with any anonymization method that conforms to Definition 1. For ease of exposition, in the rest of the paper we will adopt a specific anonymization function, namely, the *MBR (Minimum Bounding Rectangle)* generalization function [Bayardo and Agrawal 2005; Ghinita

| Age | Zipcode | Group ID |
|-----|---------|----------|
| 21 | 10000 | 1 |
| 27 | 18000 | 1 |
| 32 | 35000 | 2 |
| 32 | 35000 | 2 |
| 54 | 60000 | 3 |
| 60 | 63000 | 3 |
| 60 | 63000 | 3 |
| 60 | 63000 | 3 |

| Group ID | Disease |
|----------|---------|
| 1 | dyspepsia |
| 1 | flu |
| 2 | bronchitis |
| 2 | gastritis |
| 3 | diabetes |
| 3 | dyspepsia |
| 3 | flu |
| 3 | gastritis |

(a) The QI table            (b) The sensitive table

Table VI.   An anonymization of Table I produced from anatomy

et al. 2007; LeFevre et al. 2006a; Xiao and Tao 2007]. This function anonymizes a QI-group $G$ by replacing each $A_i^q$ ($i \in [1, d]$) value with the tightest interval that contains all $A_i^q$ values in $G$. For instance, Table III is obtained by applying the MBR function to a partition of Table I with three QI-groups {Ann, Bob}, {Cate, Don}, and {Ed, Fred, Gill, Hera}.

Let $T^*$ be the anonymization of $T$ released by the publisher. $T^*$ should satisfy $l$-*diversity*:

DEFINITION 2 ($l$-DIVERSITY [MACHANAVAJJHALA ET AL. 2007]). *A  QI-group $G$ is **$l$-diverse**, if and only if it contains at most $|G|/l$ tuples with the same sensitive value. A partition is $l$-diverse, if and only if each of its QI-groups is $l$-diverse. An anonymization is $l$-diverse, if and only if it is produced from an $l$-diverse partition.*

It is noteworthy that there exist several different definitions of $l$-diversity [Machanavajjhala et al. 2007]. For example, *entropy $l$-diversity* requires that the entropy of sensitive values in each QI-group should be at least $\ln l$; *recursive $(c, l)$-diversity* demands that, even if we remove $l - 2$ arbitrary sensitive values in a QI-group $G$, at most $c$ fraction of the remaining tuples should have the same sensitive value. Definition 2 corresponds to a simplified version of recursive $(c, l)$-diversity, and has been widely adopted previously [Ghinita et al. 2007; Xiao and Tao 2006a; Wong et al. 2006; Wong et al. 2007].

Let $\mathcal{G}$ be the anonymization algorithm adopted by the publisher. $\mathcal{G}$ can be either deterministic or randomized, but it should be an *$l$-diversity algorithm*. That is, $\mathcal{G}$ should take as input any microdata $T'$ and any positive integer $l$, and output either $\emptyset$ or an $l$-diverse anonymization of $T'$. In particular, $\mathcal{G}$ may return $\emptyset$, when no $l$-diverse anonymization exists for $T'$. For instance, given the microdata $T_1$ in Table I, no algorithm can generate a 10-diverse anonymization, since $T_1$ contains only 8 tuples.

Consider an adversary who tries to infer sensitive information from $T^*$. As demonstrated in Section 1, the adversary may employ an *external source* (e.g., a voter registration list) to identify the individuals involved in $T^*$. More formally, we define an external source $E$ as a table that contains all attributes in $T$, except $A^s$. In addition, for each tuple $t \in T$, there should exist a unique record $e \in E$, such that $t$ and $e$ coincide on all identifier and QI attributes. In other words, each individual in $T$ should appear in $E$, but not necessarily vice versa. For example, the external source $E_1$ in Table II involves all individuals in the microdata $T_1$ in Table I, but it also contains the information of *Bruce*, who does not appear in $T_1$.

In addition to $E$ and $T^*$, we also assume that the adversary knows the details of the anonymization algorithm $\mathcal{G}$ and the value of $l$ used by the publisher (in practice, $l$ can be inferred from $T^*$ [Wong et al. 2007]). We quantify the disclosure risks incurred by the publication of $T^*$ as:

DEFINITION 3 (DISCLOSURE RISK). *For any individual $o$, the* **disclosure risk** *$risk(o)$ of $o$ in $T^*$ is the tight upper-bound of the adversary's posterior belief in the event that "$o$ appears in $T$ and has a sensitive value $v$", given $T^*$, any sensitive value $v$, the external source $E$, the algorithm $\mathcal{G}$, and the value of $l$:*

$$risk(o) = \max_{v \in A^s} Pr\{o \text{ appears in } T \text{ and has a sensitive value } v \mid T^* \wedge E \wedge \mathcal{G} \wedge l\}, \quad (1)$$

*where $Pr\{X \mid Y\}$ denotes the conditional probability of event $X$ given the occurrence of event $Y$.*

### 2.2 Disclosure Risks in Anonymized Tables

Next, we present a detailed analysis of disclosure risks. Before examining $T^*$, the adversary has no information about (i) which individuals in the external source $E$ appear in $T$, and (ii) what is the sensitive value of each person. Thus, from the adversary's perspective, there exist many *possible instances* of the microdata. In particular, each instance $\hat{T}$ may involve any individuals in $E$, and each person in $\hat{T}$ can have an arbitrary sensitive value. We formally define such instances as:

DEFINITION 4 (POSSIBLE MICRODATA INSTANCE). *Given an external source $E$, a* **possible microdata instance** *based on $E$ is a microdata table $\hat{T}$ that contains a subset of the individuals in $E$, such that each of these individuals have the same QI values in $E$ and $\hat{T}$ (the sensitive value of each individual in $\hat{T}$ can be arbitrary).*

For example, given the external source in Table II, Table VII is a possible microdata instance. Note that, the microdata $T$ itself is also a possible instance. In general, possible instances may be completely different from $T$, e.g., Table I and Table VII do not even have the same cardinality. Nevertheless, it is reasonable to assume that, before inspecting $T^*$, the adversary considers each possible instance to be equally likely. This assumption is referred to as the *random worlds assumption* [Bacchus et al. 1996], and is adopted by most existing work on data anonymization[4] [Byun et al. 2006; Chen et al. 2007; Kifer and Gehrke 2006; LeFevre et al. 2006b; Li et al. 2007; Martin et al. 2007; Nergiz et al. 2007; Wang and Fung 2006; Wong et al. 2006; Wong et al. 2007; Xiao and Tao 2006b; 2007; Zhang et al. 2007; Zhang et al. 2007].

Let $S$ be the set of all possible microdata instances based on $E$. Now, consider that the adversary has obtained $T^*$, the anonymization algorithm $\mathcal{G}$, and the parameter $l$. For simplicity, assume for the moment that $\mathcal{G}$ is deterministic. The adversary can utilize the algorithm $\mathcal{G}$ to refine $S$. Specifically, s/he can apply $\mathcal{G}$ on

---

[4]Recent research [Kifer 2009] shows that, when the random worlds assumption does not hold, some of the existing anonymization methods are vulnerable to privacy attacks based on machine learning techniques. The treatment of such privacy attacks is beyond of the scope of this paper.

| Name | Age | Zipcode | Disease |
|------|-----|---------|---------|
| Bruce | 29 | 19000 | bronchitis |
| Cate | 32 | 35000 | flu |
| Fred | 60 | 63000 | dyspepsia |

Table VII.   A Possible Microdata Instance Based on Table II

**Algorithm** *Opt-Gen* $(T, l)$
1. $S_p$ = a set containing all partitions $P$ of $T$, such that $P$ and the MBR function decide an $l$-diverse global recoding generalization
2. if $S_p = \emptyset$ then return $\emptyset$
3. among all $P \in S_p$, select the one that minimizes $\sum_{G \in P} |G|^2$
4. return the generalization determined by $P$ and the MBR function

Fig. 2.   The *Opt-Gen* algorithm

each instance $\hat{T} \in S$, and inspect the output of $\mathcal{G}$. If $\hat{T}$ leads to an anonymization different from $T^*$, the adversary asserts that, $\hat{T}$ is not the real microdata $T$. Let $S'$ be the set of instances that pass the sanity check, i.e., for each $\hat{T} \in S'$, $\mathcal{G}(\hat{T}, l) = T^*$ (apparently, $T \in S'$).

The adversary then uses $S'$ to infer the sensitive information in $T$. As a special case, if an individual $o$ is associated with an $A^s$ value $v$ in all instances in $S'$, then $v$ must be the $A^s$ value of $o$ in $T$. In general, the probability that $o$ has $v$ in $T$ depends on the portion of instances in $S'$ where $o$ has $v$. We refer to the above inference approach as a *reverse engineering attack*.

EXAMPLE 2. Consider the $l$-diversity generalization algorithm *Opt-Gen*, as shown in Figure 2. In a nutshell, *Opt-Gen* employs the MBR function, and returns $l$-diverse generalizations that (i) obey global recoding, and (ii) minimize the *discernability metric* [Bayardo and Agrawal 2005]. Specifically, the discernability of a generalized table $T^*$ equals $\sum_{G \in P} |G|^2$, where $P$ is the partition that decides $T^*$.

Suppose that a publisher adopts *Opt-Gen* to anonymize the microdata $T_1$ in Table I, setting $l$ to 2. Table III illustrates the resulting generalization $T_2^*$. Assume that an adversary has the external source $E_1$ in Table II, and knows *Opt-Gen* and $l = 2$. To launch a reverse engineering attack, s/he first constructs the set $S$ of all possible microdata instances based on $E_1$ (e.g., Table VII is one instance in $S$). As a second step, the adversary invokes *Opt-Gen* on each $\hat{T} \in S$, and verifies whether the output of *Opt-Gen* is $T_2^*$. Let $S'$ be the maximal subset of $S$ such that $Opt\text{-}Gen(\hat{T}, 2) = T_2^*$ for each $\hat{T} \in S'$. In the sequel, we will show that every $\hat{T} \in S'$ must associate Ed with *gastritis*. Namely, based on $T_2^*$, $E_1$, $l = 2$, and the details of *Opt-Gen*, the adversary can infer the exact disease of Ed.

Let $G_1$, $G_2$, and $G_3$ be the first, second, and third QI-group in $T_2^*$, respectively. Any $\hat{T} \in S$, which can be generalized to $T_2^*$, must satisfy the following conditions. First, $\hat{T}$ should not involve Bruce, since his age 29 is not covered by any *Age* interval in $T_2^*$. Second, $\hat{T}$ should either (i) associate Ann with *dyspepsia* and Bob with *flu*, or (ii) conversely, associate Ann and Bob with *flu* and *dyspepsia*, respectively. This is because, Ann and Bob are the only individuals whose ages fall in the *Age* interval [21, 27] of $G_1$, while $G_1$ contains two sensitive values *dyspepsia* and *flu*. By the same reasoning, $\hat{T}$ should assign the diseases in $G_2$ ($G_3$) to Cate and Don (Ed, Fred, Gill, and Hera).

We are now ready to prove that, any possible microdata instance in $S'$ must set the sensitive value of Ed to *gastritis*. Assume, on the contrary, that this is not true in a $\hat{T}' \in S'$. Then, since Ed is in $G_3$, his disease in $\hat{T}'$ must be one of {*flu, diabetes, dyspepsia*}, i.e., the sensitive values in $G_3$ except *gastritis*. In that case, Ed's disease in $\hat{T}'$ must differ from those of Cate and Don (each of whom suffers from either *gastritis* or *bronchitis* in $\hat{T}'$). Hence, we can construct a 2-diverse QI-group $G_2' =$ {Cate, Don, Ed}. The other tuples in $\hat{T}'$ can also form two 2-diverse QI-groups $G_1' = $ {Ann, Bob}, and $G_3' = $ {Fred, Gill, Hera}.

Let $P' = \{G_1', G_2', G_3'\}$, which decides a 2-diverse global recoding generalization. Let us refer to that generalization as $T'^*$. The discernability of $T'^*$ is $2^2 + 3^2 + 3^2 = 22$, which is smaller than the discernability 24 of $T_2^*$. As *Opt-Gen* minimizes the discernability, given $\hat{T}'$ as the input, it should have output $T'^*$ instead of $T_2^*$, leading to a contradiction. In conclusion, Ed must be assigned a sensitive value *gastritis* in any $\hat{T} \in S'$.                                           □

The above discussion motivates the following proposition for computing disclosure risks.

PROPOSITION 1. *Let $o$ be any individual, $E$ be an external source, and $T^*$ be an anonymization of $T$ produced with an $l$-diversity algorithm $\mathcal{G}$ and a parameter $l$. Let $S$ be the set of possible microdata instances based on $E$. Let $S_{o,v}$ be the maximal subset of $S$, such that each instance $\hat{T} \in S_{o,v}$ includes a tuple $t$, with $t[A^{id}] = o$ and $t[A^s] = v$. Then,*

$$risk(o) = \max_{v \in A^s} \frac{\sum_{\hat{T} \in S_{o,v}} Pr\{\mathcal{G}(\hat{T}, l) = T^*\}}{\sum_{\hat{T} \in S} Pr\{\mathcal{G}(\hat{T}, l) = T^*\}}, \qquad (2)$$

*where $Pr\{\mathcal{G}(\hat{T}, l) = T^*\}$ denotes the probability that, given $\hat{T}$ and $l$, algorithm $\mathcal{G}$ outputs $T^*$.*

The proofs of all propositions, lemmas, and theorems can be found in the appendix. We are now ready to introduce the *transparent l-diversity* principle, for protecting privacy when the anonymization algorithm is "transparent" to adversaries.

DEFINITION 5 (TRANSPARENT $l$-DIVERSITY). *An anonymization $T^*$ of $T$ is* **transparently $l$-diverse** *if, given any external source, $T^*$ ensures $risk(o) \leq 1/l$ for any individual $o$. An $l$-diversity algorithm $\mathcal{G}$ is* **transparent***, if and only if given any microdata $T$ and any positive integer $l$, algorithm $\mathcal{G}$ outputs either $\emptyset$ or a transparently $l$-diverse anonymization of $T$.*

Intuitively, an $l$-diversity algorithm $\mathcal{G}$ is transparent, if and only if each output $T^*$ of $\mathcal{G}$ can be generated from a set $S$ of possible microdata instances, such that each individual $o$ is associated with a diverse set of sensitive values in different instances. As the adversary cannot decide which instance in $S$ corresponds to the input microdata, s/he would not be able to infer the exact sensitive value of $o$ from $T^*$. The fact that each instance in $S$ can lead to $T^*$ implies that the output of $\mathcal{G}$ should not be highly dependent on the sensitive value of any particular individual. For instance, the *Opt-Gen* algorithm fails in Example 2, because it outputs $T_2^*$ (in Table III) only if Ed has a sensitive value *gastritis*. In general, a transparent algorithm should anonymize data in a manner such that none of the steps of the anonymization process is uniquely decided by the sensitive value of a particular tuple. In Section 3, we will present three transparent algorithms that are developed according to the above principle.

### 2.3 Comparison with Previous Work

As explained in Section 1, [Wong et al. 2007] and [Zhang et al. 2007] are the only previous works that do not assume adversaries with no knowledge of the anonymization algorithm $\mathcal{G}$. In this section, we elaborate the solutions in [Wong et al. 2007] and [Zhang et al. 2007], and point out how they differ from our solution.

**Comparison with [Wong et al. 2007].** The privacy model in [Wong et al. 2007] assumes that (i) the anonymization algorithm $\mathcal{G}$ is deterministic, and (ii) the adversary knows whether $\mathcal{G}$ produces minimal generalization. To clarify the model, we begin by reviewing several concepts in [Wong et al. 2007].

DEFINITION 6 (CHILD PARTITION). *Let $P_1$ and $P_2$ be two partitions of $T$. $P_2$ is a **child** of $P_1$, if and only if there exist $G_1 \in P_1$ and $G_2, G_3 \in P_2$, such that (i) $G_1 = G_2 \cup G_3$, and (ii) $P_1 - \{G_1\} = P_2 - \{G_2, G_3\}$.*

Note that we can obtain a child of a partition $P$, by splitting a QI-group in $P$ into two smaller QI-groups.

DEFINITION 7 (MINIMAL GENERALIZATION). *Let $f$ be a generalization function, $P$ an $l$-diverse partition, and $T^*$ the generalization decided by $f$ and $P$. $T^*$ is a **minimal $l$-diverse generalization** under global (local) recoding, if $f$ and any child of $P$ cannot decide an $l$-diverse generalization under the same recoding.*

For example, Table III is a minimal 2-diverse generalization of Table I with respect to the MBR function and global recoding, as explained in Section 1. Given a generalization function $f$ and recoding scheme $H$, we say that an $l$-diversity algorithm is *minimal*, if it produces only minimal generalizations under $f$ and $H$. The subsequent discussion will focus on minimal algorithms $\mathcal{G}$, because the results of [Wong et al. 2007] are inapplicable to non-minimal algorithms (i.e., minimality attacks cannot be performed if $\mathcal{G}$ is non-minimal).

In a similar fashion to Definition 1, Wong et al. [2007] formulate the disclosure risks (referred to as *credibilities* in [Wong et al. 2007]) as:

DEFINITION 8 (CREDIBILITY). *Let $o$ be any individual, and $V$ be a predefined subset of the values in $A^s$. The **credibility** of $o$ in $T^*$ is the adversary's maximum posterior belief in the event that "o appears in $T$ and has a sensitive value $v$", given $T^*$, an external source $E$, generalization function $f$, recoding scheme $H$, value of $l$, and $\mathcal{G}$ being minimal:*

$$cred(o) = \max_{v \in V} Pr\{o \text{ has } v \text{ in } T \mid T^* \wedge E \wedge f \wedge H \wedge l \wedge \mathcal{G} \text{ is minimal}\}.$$

Note that the credibility model quantifies disclosure risks based only on a subset $V$ of the $A^s$ values. To facilitate the comparison between the credibility model and our privacy model, we assume $V = A^s$ in the rest of the paper.

Credibilities can be derived as:

PROPOSITION 2 [WONG ET AL. 2007]. *Let $o$, $E$, $f$, $H$, $l$ be as introduced in Definition 8. Let $S^+$ be the set including any possible microdata instance $\hat{T}$ based on $E$, such that $T^*$ is a minimal $l$-diverse generalization of $\hat{T}$ with respect to $f$ and $H$. Let $S^+_{o,v}$ be the maximal subset of $S^+$, such that in each instance in $S^+$, $o$ is associated with a sensitive value $v$. We have*

$$cred(o) = \max_{v \in A^s} |S^+_{o,v}|/|S^+|. \tag{3}$$

The following analysis will confirm the intuition that credibilities underestimate the actual privacy risks, when an adversary knows everything about $\mathcal{G}$. Towards this, let us revisit the scenario in Example 2, where the adversary can precisely find out Ed's disease with a reverse engineering attack, i.e, the disclosure risk of Ed equals the maximum value 1. In the sequel, we will show that $cred(\text{Ed}) = 1/4$.

LEMMA 1. *The Opt-Gen algorithm (in Figure 2) is a minimal algorithm.*

EXAMPLE 3. Consider the settings in Example 2, where $T = T_1$, $T^* = T^*_2$, $E = E_1$, $\mathcal{G} = Opt\text{-}Gen$, $l = 2$, $o = $ Ed. Since $Opt\text{-}Gen$ is a minimal algorithm (see Lemma 1), by Proposition 2, the credibility of Ed in $T^*_2$ is calculated as $\max_{v \in A^s} |S^+_{o,v}|/|S^+|$, where $S^+$ is the set of all possible microdata instances that have $T^*_2$ as a minimal generalization, and $S^+_{o,v}$ is the subset of instances in $S^+$ that associate Ed with a certain sensitive value $v$.

Let $\hat{T}$ be any possible microdata instance based on $E_1$. As demonstrated in Example 2, if $\hat{T}$ can be generalized to $T^*_2$, then $\hat{T}$ must not involve Bruce. Furthermore, $\hat{T}$ should assign the sensitive values in the first, second, and third QI-groups in $T^*_2$ to {Ann, Bob}, {Cate, Don}, and {Ed, Fred, Gill, Hera}, respectively. Totally, there are $2! \times 2! \times 4! = 96$ different combinations between the sensitive values and individuals. This leads to a set $S_m$ of 96 possible microdata instances. For any $v = $ *gastritis, flu, dyspepsia,* or *diabetes,* there exist 24 instances in $S_m$ that associate Ed with $v$. Since $S_m$ includes all possible microdata instances that can be generalized to $T^*_2$, we have $S^+ \subseteq S_m$.

Next, we will prove $S^+ = S_m$. For this purpose, it suffices to establish that, for any instance $\hat{T} \in S_m$, $T^*_2$ is a minimal 2-diverse generalization with respect to global recoding and the MBR function $f$. Let $G_1 = \{\text{Ann, Bob}\}$, $G_2 = \{\text{Cate, Don}\}$, $G_3 = \{\text{Ed, Fred, Gill, Hera}\}$. The partition underlying $T^*_2$ is $P_1 = \{G_1, G_2, G_3\}$. Assume, on the contrary, that $T^*_2$ is not minimal for some $\hat{T} \in S_m$. Then, there exists a partition $P_2$ of $\hat{T}$ such that (i) $P_2$ is a child of $P_1$, and (ii) $P_2$ and $f$ decide a 2-diverse global recoding generalization.

As $P_2$ is a child of $P_1$, by Definition 6, we can obtain $P_2$ from $P_1$ by splitting $G_1$, $G_2$, or $G_3$. However, it is impossible to split $G_1$ ($G_2$) into 2-diverse QI-groups, since it contains only two tuples. On the other hand, $G_3$ cannot be divided either. This is because, Fred, Gill, and Hera have identical QI values, and thus, have to be in the same QI-group (due to global recoding); meanwhile, the remaining tuple Ed itself does not make a 2-diverse QI-group. Hence, under global recoding, no child of $P_1$ can lead to a 2-diverse generalization of $\hat{T}$. It follows that $T^*_2$ is a minimal generalization of every $\hat{T} \in S_m$, i.e., $S^+ = S_m$.

Finally, since (as mentioned earlier) there exist exactly 24 instances in $S_m$ that assign the same sensitive value to Ed, $cred(\text{Ed}) = \max_{v \in A^s} |S_{o,v}^+|/|S^+| = 24/96 = 1/4$. □

Since the credibility model cannot secure privacy against an adversary who knows the anonymization algorithm, any method developed based on the model is susceptible to revere engineering attacks. To demonstrate this, we exemplify in the electronic appendix an attack against *Mask*, an anonymization approach devised in [Wong et al. 2007] under the credibility model.

**Comparison with [Zhang et al. 2007].** Zhang et al. [2007] consider the publication of microdata using deterministic algorithms that adopt global recoding. They model a global recoding generalization as a projection of the microdata into a "coarsened" multi-dimensional domain. For example, given the microdata $T_1$ in Table I, we can coarsen the *Age* domain, so that it contains only seven values: "$\leq 20$", "$[21, 27)$", "$(27, 32)$", "$32$", "$(32, 54)$", "$[54, 60]$", and "$\geq 60$". Similarly, we can define a coarsened *Zipcode* domain that has only seven values: "$< 10k$", "$[10k, 18k)$", "$(18k, 35k)$", "$35k$", "$(35k, 60k)$", "$[60k, 63k]$", and "$> 63k$". Accordingly, the global recoding generalization $T_2^*$ in Table III can be regarded as the projection of $T_1$ into the three-dimensional domain spanned by *Disease* and the coarsened *Age* and *Zipcode*. Let $C$ be the set of all coarsened multi-dimensional domains that can be constructed from the attributes in the microdata. Zhang et al. assume that the domains in $C$ can be totally ordered by their *information loss*, which measures the degree of coarseness of the domains. For example, the information loss of a domain is (i) minimized if no coarsening is applied, and (ii) maximized if every attribute is maximally coarsened.

Zhang et al. consider that the publisher adopts a deterministic generalization algorithm $\mathcal{G}$ as follows. Given a microdata $T$ and a privacy principle, $\mathcal{G}$ first examines the multi-dimensional domains in $C$ in ascending order of their information loss. For each domain $D^*$, $\mathcal{G}$ projects $T$ into $D^*$, and checks whether the resulting generalization satisfies the given privacy principle. If the principle is satisfied, $\mathcal{G}$ returns the generalization and terminates; otherwise, $\mathcal{G}$ moves on to the next domain in $C$. In other words, $\mathcal{G}$ always outputs the first generalization that conforms to the adopted privacy principle. Alternatively, $\mathcal{G}$ may also traverse $C$ in descending order of information loss, and returns the last generalization on which the given principle is satisfied. The adversary is assumed to (i) have an external source $E$ that contains only the individuals in the microdata, and (ii) know the privacy principle as well as the order in which $\mathcal{G}$ traverses $C$.

Under the above problem setting, Zhang et al. present a theoretical study on how $\mathcal{G}$ should be designed to prevent the adversary from inferring private information. Let $n_p$ be the total number of possible microdata instances based on $E$. Zhang et al. first prove that it is NP-hard (with respect to $n_p$) to compute a generalization that both ensures privacy and incurs the minimum information loss. After that, they investigate three special cases of the problem by imposing various constraints on $C$ and the privacy principle. For each case, they show that the optimal generalization can be computed in time polynomial in $n_p$ and the size of $C$. Finally, they propose a generalization algorithm that ensures *entropy l-diversity* (see Section 2.1), and prove that its time complexity is polynomial in $|C|$ and independent of $n_p$. Note that, in practice, both $|C|$ and $n_p$ are usually exponential in the number $n$ of tuples in the microdata.

Compared with the solution in [Wong et al. 2007], Zhang et al.'s techniques achieve a higher level of privacy protection, as they can guard against an adversary who has full knowledge of the anonymization algorithm. Nevertheless, Zhang et al.'s work has the following limitations. First, the privacy model in [Zhang et al. 2007] is restricted to a particular type of deterministic algorithms that adopt global recoding. Consequently, the model cannot be used to evaluate the privacy guarantee of any existing anonymization algorithm that is randomized or local-recoding-based, nor does it support the development of new anonymization approaches of those kinds. Second, all algorithms proposed in [Zhang et al. 2007] have time complexities exponential in the number $n$ of tuples in the microdata, and there is no experimental evaluation included in [Zhang et al. 2007] to demonstrate the effectiveness or efficiency of the algorithms. This leaves open the question of whether or not the algorithms in [Zhang et al. 2007] are applicable in practice.

Our work remedies the deficiencies of [Zhang et al. 2007]. In particular, our privacy model captures all (deterministic or randomized) anonymization algorithms that adopt generalization or anatomy. This general model enables us to design three transparent anonymization algorithms, all of which fall beyond Zhang et al.'s model as they rely on random choices and/or local recoding. In addition, as will be shown in Section 3, our algorithms run in $O(n^2 \log n)$ time, which significantly improves over the exponential time complexities of Zhang et al.'s techniques. Finally, we will present in Section 4 an extensive experimental study that demonstrates the practical performance of our algorithms in terms of data utility and computation time.

## 3. ACHIEVING TRANSPARENT $L$-DIVERSITY

Equipped with the analytical model in Section 2, our next step is to develop transparent anonymization algorithms for $l$-diversity. Ideally, an algorithm should produce anonymizations with minimum information loss, according to a certain *penalty metric h*. Specifically, $h$ is a function that, given a QI-group $G$, calculates a penalty $h(G)$ based on the tuples in $G$. Given $h$, the information loss of an anonymization $T^*$ is computed as $\sum_{G \in P} h(G)$, where $P$ is the partition underlying $T^*$. For example, the discernibility metric deployed in Example 2 corresponds to a function $h_d$ such that $h_d(G) = |G|^2$ for any QI-group $G$.

In the following, we will elaborate three transparent algorithms, each of which can be combined with any penalty metric $h$, as long as the metric (i) does not rely on the sensitive values in the input QI-group, and (ii) is *superadditive*, i.e., $h(G_1 \cup G_2) \geq h(G_1) + h(G_2)$ holds for any disjoint QI-groups $G_1$ and $G_2$. For our discussion, we use the *perimeter function $h_p$* [Ghinita et al. 2007; Iyengar 2002] as a representative:

$$h_p(G) = |G| \cdot \sum_{i=1}^{d} \frac{\max_{t \in G}\left\{t[A_i^q]\right\} - \min_{t \in G}\left\{t[A_i^q]\right\}}{\max\left\{A_i^q\right\} - \min\left\{A_i^q\right\}}. \tag{4}$$

Given a set $S_G$ of QI-groups, we refer to $\sum_{G \in S_G} h_p(G)$ as the *perimeter* of $S_G$.

### 3.1 The *Tailor* Algorithm

3.1.1 *Algorithm Description.* This section presents a transparent algorithm, *Tailor*, which produces anonymized tables in a manner similar to the construction of kd-trees [Friedman et al. 1977]. *Tailor* requires the microdata $T$ to be *l-eligible*. That is, at most $|T|/l$ tuples in $T$ have the same sensitive value. If $T$ is not $l$-eligible, *Tailor* returns $\emptyset$, since no $l$-diverse anonymization of $T$ exists [Machanavajjhala et al. 2007].

Given an $l$-eligible $T$, *Tailor* first creates a partition $P$ with only one QI-group $G_0$, which includes all tuples in $T$. As a second step, *Tailor* tries to split $G_0$ into two $l$-diverse subsets $G_1$ and $G_2$ subject to certain constraints to be clarified later. If splitting is possible, *Tailor* removes $G_0$ from $P$, and inserts $G_1$ and $G_2$ in $P$. This decreases the perimeter of $P$. After that, *Tailor* recursively splits a QI-group in $P$, until no QI-group can be divided further, i.e., the perimeter of $P$ has reached a local minimum. Then, *Tailor* terminates, and outputs the anonymization decided by $P$ and an anonymization function (e.g., the MBR function).

Whenever *Tailor* divides a QI-group $G$ into subsets $G_a$ and $G_b$, $\{G_a, G_b\}$ must be an *l-cut*:

DEFINITION 9 ($l$-CUT). *Let $G$ be a QI-group, $l$ be a positive integer, and $c$ be the maximum number of tuples in $G$ with the same sensitive value. An **l-cut** of $G$ on $A_i^q$ ($i \in [1, d]$) is an ordered set $\{G_a, G_b\}$ of QI-groups, such that:*

(1) *$G_a \cup G_b = G$, and $G_a \cap G_b = \emptyset$.*
(2) *$|G_a| \geq l \cdot c$ and $|G_b| \geq l \cdot c$.*
(3) *For any $t_a \in G_a$ and $t_b \in G_b$, either (i) $t_a[A_i^q] < t_b[A_i^q]$, or (ii) $t_a[A_i^q] = t_b[A_i^q]$ and $t_a[A^{id}] < t_b[A^{id}]$.*

*The **perimeter** of the l-cut is the total perimeter of $G_a$ and $G_b$.*

Condition 2 in Definition 9 implies that $G$ (on which the $l$-cut is performed) is $2l$-diverse. Condition 3 requires, intuitively, that all tuples in $G_a$ must precede those in $G_b$, along the dimension $A_i^q$ on which $G$ is divided.

Interestingly, *as long as $G$ is $2l$-diverse, there exists at least one $l$-cut on any QI-attribute $A_i^q$ ($i \in [1, d]$).* Such a cut can be found as follows. First, we sort the tuples in $G$ in ascending order of their $A_i^q$ values. In case two tuples have the same value on $A_i^q$, the tuple with a smaller identifier precedes the other. Then, we create $G_a$ by including the first $k$ tuples in the sorted sequence (for any $k \in [l \cdot c, |G| - l \cdot c]$), and construct $G_b$ using the remaining tuples.

The above strategy yields totally $d \cdot (|G| + 1 - 2l \cdot c)$ different $l$-cuts. Among them, *Tailor* always selects the *canonical* one:

DEFINITION 10 (CANONICAL $l$-CUT). *The **canonical l-cut** of a QI-group $G$ is the l-cut with the smallest perimeter. In case multiple l-cuts have the smallest parameter, the canonical l-cut $\{G_a, G_b\}$ is uniquely decided as follows. Assume $\{G_a, G_b\}$ is on dimension $A_i^q$ ($i \in [1, d]$); then:*

(1) *No l-cut on any $A_j^q$ ($j < i$) has the same perimeter as $\{G_a, G_b\}$.*
(2) *For any l-cut $\{G'_a, G'_b\}$ on $A_i^q$, if $\{G'_a, G'_b\}$ and $\{G_a, G_b\}$ have the same perimeter, it must hold that $|G_a| < |G'_a|$.*

**Algorithm** *Tailor* $(T, l)$
1.  if $T$ is not $l$-eligible then return $\emptyset$
2.  $G_0 =$ a QI-group containing all tuples in $T$, and $P = \{G_0\}$
3.  while there exists a $2l$-diverse QI-group $G$ in $P$
4.      $\{G_a, G_b\} =$ the canonical $l$-cut of $G$
5.      $P = P - \{G\} + \{G_a, G_b\}$
6.  return the anonymization decided by $P$ and an anonymization fucntion

Fig. 3.   The *Tailor* algorithm

| Name | Age | Zipcode | Disease |
|------|-----|---------|---------|
| Ann  | 21  | 10000   | dyspepsia |
| Bob  | 27  | 18000   | flu |
| Cate | 32  | 35000   | gastritis |
| Don  | 32  | 35000   | gastritis |
| Ed   | 54  | 60000   | flu |
| Fred | 60  | 63000   | bronchitis |
| Gill | 60  | 63000   | dyspepsia |
| Hera | 60  | 63000   | diabetes |

Table VIII.   Microdata $T_5$

| Age | Zipcode | Disease |
|-----|---------|---------|
| [21, 32] | [10k, 35k] | dyspepsia |
| [21, 32] | [10k, 35k] | flu |
| [21, 32] | [10k, 35k] | gastritis |
| [21, 32] | [10k, 35k] | gastritis |
| [54, 60] | [60k, 63k] | flu |
| [54, 60] | [60k, 63k] | bronchitis |
| 60 | 63000 | dyspepsia |
| 60 | 63000 | diabetes |

Table IX.   Generalization $T_6^*$

Note that the canonical $l$-cut of a QI-group $G$ is determined by (i) the identifiers and QI values in $G$, as well as (ii) the maximum number $c$ of tuples in $G$ with the same sensitive value – all of this information is independent of the concrete sensitive value of any particular tuple. This property is the key to ensuring transparent $l$-diversity, as will be discussed in Section 3.1.2.

Figure 3 shows the pseudo-code of *Tailor*. We demonstrate the algorithm with an example, assuming that the MBR function is adopted.

EXAMPLE 4. Let us use *Tailor* to obtain a transparently 2-diverse generalization of the microdata $T_5$ in Table VIII (i.e., $T = T_5$ and $l = 2$). *Tailor* first verifies that $T_5$ is 2-eligible (Line 1 in Figure 3), and then initializes a partition $P = \{G_0\}$, where $G_0 = T_5$ (Line 2). The subsequent execution of *Tailor* is in iterations (Lines 3-5). In each iteration, *Tailor* looks for a 4 ($= 2l$) diverse QI-group $G$ in $P$ (Line 3). If $G$ does not exist, *Tailor* terminates, and returns the generalization decided by $P$ (Line 6). Otherwise, *Tailor* splits $G$ using its canonical $l$-cut (Lines 4-5), and replaces $G$ with the new QI-groups.

Specifically, in the first iteration, the only QI-group $G_0$ in $P$ is 4-diverse, and hence, is chosen to be split. *Tailor* identifies $c = 2$, which, as in Definition 9, is the largest number of tuples in $G_0$ having the same sensitive value. Then, *Tailor* proceeds to find the canonical 2-cut of $G_0$. For this purpose, it needs to obtain the best 2-cut (with the smallest perimeter) along every dimension. Dealing with *Age* first, *Tailor* sorts the tuples in $G_0$ by their *Age* values, and tries all possibilities of dividing the sorted list into two parts, each with at least 4 ($= 2c$) tuples (required by condition 2 in Definition 9). There is only possibility: $\{G_2, G_3\}$, where $G_2 = \{$Ann, Bob, Cate, Don$\}$, and $G_3 = \{$Ed, Fred, Gill, Hera$\}$. Hence, $\{G_2, G_3\}$ is the best 2-cut on *Age*. Switching to dimension *Zipcode*, *Tailor* sorts the tuples in $G_0$ by their *Zipcode* values, and again, attempts all division possibilities. Again, $\{G_2, G_3\}$ is the only possibility, and hence, is also the best 2-cut on *Zipcode*. Hence, $\{G_2, G_3\}$ is the canonical 2-cut. *Tailor* thus replaces $G_0$ with $G_2$ and $G_3$ in $P$.

In the second iteration, $P = \{G_2, G_3\}$. As $G_2$ is not 4-diverse, it cannot be split. But $G_3$ is 4-diverse, and thus, is split using its canonical cut $\{G_4, G_5\}$, where $G_4 = \{\text{Ed, Fred}\}$ and $G_5 = \{\text{Gill, Hera}\}$. Now, $P$ becomes $\{G_2, G_4, G_5\}$. Since no QI-group is 4-diverse, *Tailor* returns the generalization $T_6^*$ determined by $P$, as shown in Table IX. □

*Tailor* is deterministic, i.e., for any $T$, $l$, and $T^*$, $Pr\{Tailor(T, l) = T^*\}$ (see Proposition 1) equals either 0 or 1. In addition, *Tailor* has an $O(n^2 \log n)$ time complexity, where $n$ is the number of tuples in $T$. This follows from the facts that (i) *Tailor* performs at most $n/l$ $l$-cuts on $T$, and (ii) each $l$-cut takes $O(n \log n)$ time.

3.1.2 *Proof of Transparent l-Diversity.* In this section, we will prove that *Tailor* ensures transparent $l$-diversity. The core of our proof is an analysis on the set $S$ of all possible microdata instances based on the adversary's external source $E$. We will show that $S$ can be divided into several subsets, such that for each subset $S_{sub}$, (i) all instances in $S_{sub}$ can be transformed to the same anonymization $T^*$ by *Tailor*, and (ii) each individual in $E$ is assigned many different sensitive values in different instances in $S_{sub}$. Intuitively, when the adversary observes $T^*$, s/he would not be able to infer which instance in $S_{sub}$ is the real microdata, and hence, the sensitive value of each individual can be concealed.

More specifically, our analysis exploits the *isomorphism* between partitions. We say that a partition $P_1$ of a possible microdata instance is isomorphic to a partition $P_2$ of another instance, if and only if each QI-group in $P_1$ is isomorphic to a QI-group in $P_2$, and vice versa (see Section 2.1 for the definition of QI-group isomorphism).

EXAMPLE 5. Consider the partition $P$ of $T_5$ (in Table VIII) generated by *Tailor* in Example 4. $P$ contains three QI-groups, namely, $G_2 = \{\text{Ann, Bob, Cate, Don}\}$, $G_4 = \{\text{Ed, Fred}\}$, and $G_5 = \{\text{Gill, Hera}\}$. The sensitive values of Ed and Fred are *flu* and *bronchitis*, respectively. Suppose that we modify the two tuples in $G_4$ by swapping their *Disease* values, such that Ed has *bronchitis* and Fred has *flu*. The resulting QI-group $G_4'$ is isomorphic to $G_4$, while the partition $P' = \{G_2, G_4', G_5\}$ isomorphic to $P$. Note that $P'$ is *not* a partition of $T_5$, but is in fact a partition of the microdata $T_3$ in Table IV (this will be useful in demonstrating Lemma 3 later). □

Recall that, for any anonymization function $f$ and any two isomorphic QI-groups $G_1$ and $G_2$, we have $f(G_1) = f(G_2)$ (see Definition 1). Therefore, once $f$ is fixed, isomorphic partitions always lead to the same anonymization. For instance, consider the partitions $P$ and $P'$ in Example 5. Notice that, $P'$ and the MBR function decide $T_6^*$ (in Table IX), which is determined by $P$ and the MBR function as well. In addition, isomorphic QI-groups have a crucial property:

LEMMA 2. *Let $G$ and $G'$ be two isomorphic QI-groups, and $\{G_1, G_2\}$ ($\{G_1', G_2'\}$) be the canonical l-cut of $G$ ($G'$). Then, $G_1$ and $G_1'$ ($G_2$ and $G_2'$) must involve the same set of individuals.*

The above lemma is fairly intuitive. Recall that, the canonical $l$-cut of a QI-

group $G$ depends *only* on the identifiers and QI values in $G$, and is independent of the sensitive values. Since isomorphic QI-groups contain equivalent identifiers and QI values, their canonical $l$-cuts divide them in the same way, and thus Lemma 2 holds. Based on Lemma 2, we derive the following result, which shows an important characteristic of *Tailor*.

LEMMA 3. *Let $T_1$ be a microdata table, $l$ be an integer, and $T^* = Tailor(T_1, l)$. Let $P_1$ be the partition of $T_1$ that decides $T^*$, $P_2$ be a partition isomorphic to $P_1$, and $T_2 = \bigcup_{G \in P_2} G$. Then, $Tailor(T_2, l) = T^*$, and $P_2$ is the partition of $T_2$ that decides $T^*$.*

For instance, consider the microdata $T_3$, $T_5$ and the partitions $P$, $P'$ in Example 5. We have shown in Example 4 that $Tailor(T_5, 2) = T_6^*$, where $T_6^*$ is decided by $P$. Recall that $P'$ is isomorphic to $P$, and $T_3 = \bigcup_{G \in P'} G$. According to Lemma 3, we have $Tailor(T_3, 2) = T_6^*$, i.e., given $l = 2$, *Tailor* transforms both $T_3$ and $T_5$ into $T_6^*$.

The following theorem shows a sufficient condition for transparent $l$-diversity.

THEOREM 1. *An $l$-diversity algorithm $\mathcal{G}$ is transparent if it satisfies the following condition: For any microdata $T_1$ such that $\mathcal{G}(T_1, l) = T^*$, we have $\mathcal{G}(T_2, l) = T^*$ for a microdata table $T_2$, if $T_2$ has a partition isomorphic to the partition of $T_1$ that decides $T^*$.*

By Lemma 3, *Tailor* satisfies the sufficient condition in Theorem 1, which proves that *Tailor* is a transparent algorithm.

### 3.2 The *Ace* Algorithm

This section discusses another algorithm, *Ace* (a̲ssign and sli̲ce), which first appeared in [Xiao and Tao 2007] as part of a solution to anonymizing dynamic datasets. Here, we present non-trivial proofs on the privacy guarantee of *Ace* against adversaries who have full knowledge of the algorithm.

#### 3.2.1 *Algorithm Description.* Let us first introduce several concepts. Given a QI-group $B$, we define the *signature* of $B$ as the set of sensitive values in $B$. A *column* of $B$ refers to a maximal set of tuples in $B$ with the same sensitive value. $B$ is a *bucket*, if all of its columns contain an equal number of tuples. A partition $U$ is a *bucket partition*, if each QI-group in $U$ is a bucket.

For example, consider a QI-group $B_1$ of the microdata $T_5$ in Table VIII, where $B_1 = \{\text{Ann, Bob, Ed, Gill}\}$. The signature of $B_1$ is $\{\text{dyspepsia, flu}\}$. $B_1$ contains two columns, $L_1 = \{\text{Ann, Gill}\}$ and $L_2 = \{\text{Bob, Ed}\}$, where all tuples in $L_1$ ($L_2$) have sensitive value *dyspepsia* (*flu*). Since $|L_1| = |L_2|$, $B_1$ is a bucket. Let $B_2$ and $B_3$ be another two QI-groups of $T_5$, such that $B_2 = \{\text{Don, Fred}\}$ and $B_3 = \{\text{Cate, Hera}\}$. It can be verified that, $B_2$ and $B_3$ are also buckets. Therefore, the partition $U_1 = \{B_1, B_2, B_3\}$ is a bucket partition of $T_5$. Figure 4 illustrates $U_1$.

Apparently, $U_1$ is 2-diverse. Suppose that we divide $B_1$ into two smaller buckets, $B_4 = \{\text{Ann, Bob}\}$ and $B_5 = \{\text{Gill, Ed}\}$, both having the same signature as $B_1$. The partition $U_1' = \{B_2, B_3, B_4, B_5\}$ is also 2-diverse, and has a lower perimeter than $U_1$. In general, given any $l$-diverse bucket partition $U$, we may reduce its perimeter by splitting the buckets in $U$, without violating $l$-diversity. This strategy is adopted

| Ann | Bob |
|-----|-----|
| Gill | Ed |
| *dyspepsia* | *flu* |

$B_1$

| Don | Fred |
|-----|------|
| *gastritis* | *bronchitis* |

$B_2$

| Hera | Cate |
|------|------|
| *diabetes* | *gastritis* |

$B_3$

Fig. 4. Bucket Partition $U_1$

by *Ace*. In particular, whenever *Ace* splits a bucket $B$, the resulting sub-buckets always constitute a *division* of $B$, as defined below:

DEFINITION 11 (DIVISION). *A* **division** *of a bucket $B$ on $A_i^q$ $(i \in [1, d])$ is an ordered set $\{B_a, B_b\}$ of buckets, such that:*

(1) *$B_a \cup B_b = B$, and $B_a \cap B_b = \emptyset$.*

(2) *$B$, $B_a$ and $B_b$ have an identical signature.*

(3) *For any two tuples $t_a \in B_a$ and $t_b \in B_b$ with the same sensitive value, we have either (i) $t_a[A_i^q] < t_b[A_i^q]$, or (ii) $t_a[A_i^q] = t_b[A_i^q]$ and $t_a[A^{id}] < t_b[A^{id}]$.*

*The* **perimeter** *of the division equals the perimeter of $\{B_a, B_b\}$. A bucket is* **divisible***, if each of its columns has at least two tuples.*

Given a bucket $B$ with $x$ columns, we can obtain a division $\{B_a, B_b\}$ of $B$ on $A_i^q$ $(i \in [1, d])$ as follows. First, we sort the tuples in each column of $B$ in ascending order of their $A_i^q$ values. Whenever two tuples have an identical value on $A_i^q$, the tuple with a smaller identifier precedes the other. This results in $x$ sorted sequences. To construct $B_a$, we can remove an equal number of tuples from the top of each sequence, and insert them into $B_a$. After that, $B_b$ can be formed using the remaining tuples.

A bucket may have multiple divisions. In a way similar to canonical *l*-cuts, we formulate *canonical division* as:

DEFINITION 12 (CANONICAL DIVISION). *The* **canonical division** *of a bucket $B$ is the division with the smallest perimeter. In case multiple divisions have the smallest perimeter, the canonical division $\{B_a, B_b\}$ is uniquely decided as follows. Assume $\{B_a, B_b\}$ is on dimension $A_i^q$ $(i \in [1, d])$; then:*

(1) *No division on any $A_j^q$ $(j < i)$ has the same perimeter as $\{B_a, B_b\}$.*

(2) *For any division $\{B_a', B_b'\}$ on $A_i^q$, if $\{B_a, B_b\}$ and $\{B_a', B_b'\}$ have the same perimeter, it must hold that $|B_a| < |B_a'|$.*

As with canonical *l*-cuts, the canonical division of a bucket $B$ is irrelevant to the sensitive values in $B$. Instead, it is decided only by the identifiers and QI values in each column. In Section 3.2.2, we will exploit this property to prove that *Ace* is transparent.

Figure 5 illustrates the pseudo-code of *Ace*. Given a microdata $T$ and a positive integer $l$, *Ace* first verifies whether $T$ is *l*-eligible. After that, it invokes a subroutine *Assign* (in Figure 6) to construct an *l*-diverse bucket partition $U$ of $T$. Next, *Ace* employs the *Slice* algorithm (in Figure 8) to split the buckets in $U$, and obtains a refined partition $U'$ of $T$. In particular, the construction of $U$ is performed without inspecting the QI values of the tuples, while the split of each bucket in $U$ is based

**Algorithm** *Ace* $(T, l)$
1.  if $T$ is not $l$-eligible then return $\emptyset$
2.  $U = Assign(T, l)$
3.  $U' = Slice(U)$
4.  return the generalization decided by $U'$ and an anonymization function

Fig. 5.   The *Ace* algorithm

on canonical divisions, which are independent of the sensitive value in each column of the bucket. In other words, *Assign* and *Slice* do not rely on the correlations between the QI and sensitive values, which helps achieve transparent *l*-diversity. Finally, *Ace* returns the generalization decided by $U'$. In the following, we explain the details of *Ace* with an example, assuming that the MBR function is adopted.

EXAMPLE 6. Assume that we apply *Ace* on the microdata $T_5$ in Table VIII, with $l = 2$. *Ace* begins by checking whether $T_5$ is *l*-eligible. Since $T_5$ is 2-eligible, *Ace* invokes *Assign* to construct a bucket partition $U$ of $T_5$.

*Assign* first sets $U = \emptyset$, and creates a set $S_t$ containing all tuples in $T_5$ (Lines 1-3 in Figure 6). After that, *Assign* iteratively removes tuples from $S_t$ to construct buckets in $U$, until $S_t$ is empty (Lines 4-13). In each iteration, *Assign* first counts the frequency of each sensitive value in $S_t$ (Lines 5-6), and then builds a bucket $B$, such that (i) the signature of $B$ consists of the $\beta$ most frequent sensitive values in $S_t$, and (ii) each column of $B$ contains $\alpha$ tuples in $S_t$. The values of $\alpha$ and $\beta$ are decided in Lines 7-10, which, as explained in [Xiao and Tao 2007], guarantee that (i) $\beta \geq l$, (ii) $\alpha \geq 1$, and (iii) *Assign* always terminates[5]. For our discussion, it suffices to know that, $\alpha$ and $\beta$ depend only on the size of $S_t$ and the sensitive values in $S_t$. Since $\beta \geq l$, any bucket $B$ created by *Assign* is *l*-diverse.

In the first iteration, $S_t = T_5$, and $\alpha = \beta = 2$ (calculated by Lines 7-10). Figure 7(a) illustrates the tuples in $S_t$. *Assign* first creates a bucket $B_1$ whose signature consists of the $\beta = 2$ most frequent sensitive values in $S_t$. As shown in Figure 7(a), there exist three sensitive values in $S_t$, *dyspepsia*, *flu*, and *gastritis*, that have the same highest frequency. To pick two of the three diseases, *Assign* resorts to a total ordering. In general, any total ordering works, but for our illustration, we use the alphabetic order, in which case the signature of $B_1$ is selected as {*dyspepsia*, *flu*}. Next, for each disease in the signature, *Assign* adds $\alpha = 2$ tuples to $B_1$. As a consequence, $B_1$ contains four tuples {Ann, Bob, Gill, Ed}, as illustrated in Figure 4. The tuples in $B_1$ are then removed from $S_t$.

In the second iteration, $S_t$ contains four tuples, as shown in Figure 7(b). This time, $\alpha = 1$ and $\beta = 2$. Hence, *Assign* yields a bucket $B_2$ with signature {*gastritis*, *bronchitis*} (*gastritis* is picked as it has the highest frequency in $S_t$; *bronchitis* is chosen because it alphabetically ranks before *diabetes*). Accordingly, *Assign* inserts two tuples into $B_2$: one with a sensitive value *gastritis*, and the other one with *bronchitis*. As there are two tuples having *bronchitis*, the one to appear in $B_2$ is randomly chosen; suppose that we pick Don. This leads to $B_2 = \{Don, Fred\}$,

---
[5]Intuitively, *Assign* always terminates, because (i) each iteration of *Assign* removes $\alpha \cdot \beta > 0$ tuples from $S_t$, and hence, (ii) $S_t$ will become empty after a certain number of iterations, in which case *Assign* stops by returning the bucket partition $U$ it constructs (see Lines 3 and 13 in Figure 6).

**Algorithm** *Assign* $(T, l)$
1.  initialize a partition $U = \emptyset$
2.  $w$ = the number of distinct $A^s$ value in $T$
3.  $S_t = T$
    /\* The tuples in $S_t$ will be iteratively removed to construct buckets in $U$ \*/
4.  while $S_t \neq \emptyset$
    /\* Lines 5-12 create a new bucket in $U$ using tuples from $S_t$ \*/
5.      let $v_i$ $(i \in [1, w])$ be the $i$-th most frequent $A^s$ value in the current $S_t$
    /\* Ties are resolved by a total ordering on $A^s$ (see Example 6) \*/
6.      let $n_i$ $(i \in [1, w])$ be the number of tuples in $S_t$ with sensitive value $v_i$
7.      $\beta = l$
    /\* the new bucket's signature will contain the $\beta$ most frequent $A^s$ values in $S_t$ \*/
8.      $\alpha$ = the largest positive integer satisfying three inequalities:
    $$\alpha \leq n_\beta, \quad n_1 - \alpha \leq \frac{|S_t| - \alpha \cdot \beta}{l}, \quad \text{and } n_{\beta+1} \leq \frac{|S_t| - \alpha \cdot \beta}{l}$$
    /\* the new bucket will contain $\alpha$ tuples for each sensitive value in its signature \*/
8.      if $\alpha$ does not exist
9.          $\beta = \beta + 1$; goto Line 7
10.     create in $U$ a bucket $B$ with a signature $\{v_1, ..., v_\beta\}$
11.     for $i = 1$ to $\beta$
12.         from $S_t$, randomly remove $\alpha$ tuples whose sensitive values equal $v_i$, and insert those tuples into $B$
13. return $U$

Fig. 6.   The *Assign* algorithm

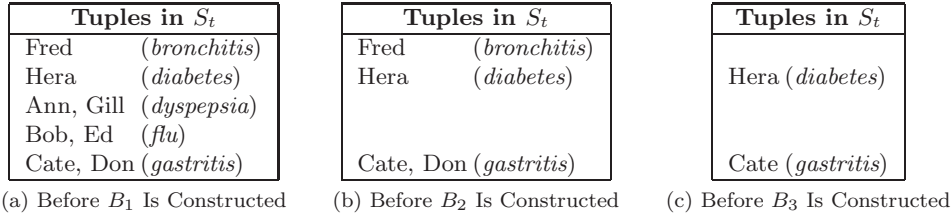

(a) Before $B_1$ Is Constructed     (b) Before $B_2$ Is Constructed     (c) Before $B_3$ Is Constructed

Fig. 7.   Changes in $S_t$ During the Execution of *Assign* in Example 6

**Algorithm** *Slice* $(U)$
1.  while there exists a divisible bucket $B$ in $U$
2.      $\{B_a, B_b\}$ = the canonical division of $B$
3.      $U = U - \{B\} + \{B_a, B_b\}$
4.  return $U$

Fig. 8.   The *Slice* algorithm

as illustrated in Figure 4. Don and Fred are then evicted from $S_t$, as shown in Figure 7(c).

Similarly, the third iteration constructs a bucket $B_3 = \{\text{Hera, Cate}\}$ (see Figure 4). Then, $S_t$ becomes empty, and hence, *Assign* terminates with a bucket partition $U = \{B_1, B_2, B_3\}$.

As the second step, *Ace* applies *Slice* to divide the buckets in $U$ into smaller QI-groups. *Slice* also runs in iterations. In each iteration, it first identifies a divisible bucket $B$ in $U$ (Line 1 in Figure 8), and then, splits $B$ using its canonical division $\{B_a, B_b\}$. This is repeated until no bucket in $U$ is divisible.

| Age | Zipcode | Disease |
|-----|---------|---------|
| [21, 27] | [10k, 18k] | dyspepsia |
| [21, 27] | [10k, 18k] | flu |
| [54, 60] | [60k, 63k] | dyspepsia |
| [54, 60] | [60k, 63k] | flu |
| [32, 60] | [35k, 63k] | gastritis |
| [32, 60] | [35k, 63k] | bronchitis |
| [32, 60] | [35k, 63k] | diabetes |
| [32, 60] | [35k, 63k] | gastritis |

Table X.  Generalization $T_7^*$

| Name | Age | Zipcode | Disease |
|------|-----|---------|---------|
| Ann | 21 | 10000 | flu |
| Bob | 27 | 18000 | dyspepsia |
| Cate | 32 | 35000 | gastritis |
| Don | 32 | 35000 | gastritis |
| Ed | 54 | 60000 | dyspepsia |
| Fred | 60 | 63000 | bronchitis |
| Gill | 60 | 63000 | flu |
| Hera | 60 | 63000 | diabetes |

Table XI.  Microdata $T_8$

In our example, the input to *Slice* is the bucket partition $U = \{B_1, B_2, B_3\}$ in Figure 4. $B_1$ is the only divisible bucket. To determine the canonical division of $B_1$, *Slice* finds the best division on each dimension (with the lowest perimeter). It turns out that, on both dimensions *Age* and *Zipcode*, the best division is $\{B_4, B_5\}$, where $B_4 = \{\text{Ann, Bob}\}$ and $B_5 = \{\text{Gill, Ed}\}$. Thus, $\{B_4, B_5\}$ becomes the canonical division. Therefore, *Slice* removes $B_1$ from $U$, and inserts $B_4$ and $B_5$ instead, leading to $U = \{B_2, B_3, B_4, B_5\}$. As no bucket in $U$ is divisible, *Slice* returns $U$ to *Ace*. Finally, *Ace* reports the generalization $T_7^*$ (in Table X) decided by $U$.     □

*Ace* is a randomized algorithm, due to the randomness in its component *Assign*. Furthermore, *Ace* has an $O(n^2 \log n)$ time complexity, where $n$ is the number of tuples in $T$. To understand this, observe that *Assign* runs in $O(n)$ time (we regard the number of distinct $A^s$ values in $T$ as a constant). On the other hand, *Slice* has an $O(n^2 \log n)$ time complexity, since (i) each bucket $B$ generated from *Assign* is divided by *Slice* exactly $|B|/l$ times, (ii) each division of $B$ incurs $O(|B| \log |B|)$ overhead, and (iii) the sizes of all buckets add up to $n$. Since *Ace* is a composition of *Assign* and *Slice*, its time complexity is $O(n^2 \log n)$.

### 3.2.2  *Proof of Transparent l-Diversity.*

This section proves that *Ace* achieves transparent $l$-diversity. Our analysis utilizes a crucial concept, the *symmetry* between buckets.

DEFINITION 13 (SYMMETRY).  *Two buckets $B_1$ and $B_2$ are **symmetric**, if and only if (i) $B_1$ and $B_2$ have the same signature, and (ii) for any column $L_1 \subseteq B_1$, there exists a column $L_2 \subseteq B_2$, such that $L_1$ and $L_2$ involve the same set of individuals. Two bucket partitions $U_1$ and $U_2$ are symmetric, if each bucket in $U_1$ is symmetric to a bucket in $U_2$, and vice versa.*

Consider, for example, the bucket partition $U_1$ in Figure 4. Bucket $B_1 \in U_1$ contains two columns $L_1 = \{\text{Ann, Gill}\}$ and $L_2 = \{\text{Bob, Ed}\}$. Suppose that we exchange the sensitive values between $L_1$ and $L_2$, by setting the sensitive values of the tuples in $L_1$ ($L_2$) to *flu* (*dyspepsia*). Then, we obtain a bucket $B_1'$ symmetric to $B_1$, as shown in Figure 9. The bucket partition $U_2 = \{B_1', B_2, B_3\}$ is symmetric to $U_1$. In general, we can obtain any symmetric counterpart of a bucket $B$, by swapping the sensitive values between different columns of $B$.

Fig. 9. Bucket partition $U_2$

Interestingly, the canonical division of a symmetric bucket always results in symmetric sub-buckets:

LEMMA 4. *Let $B$ and $B'$ be two symmetric buckets, and $\{B_1, B_2\}$ ($\{B_1', B_2'\}$) be the canonical division of $B$ ($B'$). Then, $B_1$ and $B_1'$ ($B_2$ and $B_2'$) are symmetric.*

The rationale behind Lemma 4 is similar to that of Lemma 2. Specifically, since $B$ and $B'$ are symmetric, each column $L$ in $B$ can be mapped to a column $L'$ in $B'$, such that $L$ and $L'$ involve an identical set of identifiers and QI values. Recall that the canonical division of a bucket depends only on identifiers and QI-values, and is irrelevant to sensitive values. Hence, the canonical division of $B$ has the same effect as that of $B'$, thus establishing Lemma 4. The lemma naturally leads to the following result.

LEMMA 5. *Let $U_1$ and $U_2$ be two symmetric bucket partitions. Let $U_1' = Slice(U_1)$ and $U_2' = Slice(U_2)$. Then, $U_1'$ and $U_2'$ are symmetric.*

*Assign* also has an interesting property related to symmetric buckets:

LEMMA 6. *Let $T_1$ be a microdata table, $l$ an integer, and $U_1$ a possible output of $Assign(T_1, l)$. Let $U_2$ be a bucket partition symmetric to $U_1$, and $T_2 = \bigcup_{B \in U_2} B$. Then, $Pr\{Assign(T_1, l) = U_1\} = Pr\{Assign(T_2, l) = U_2\}$.*

For instance, consider the symmetric bucket partitions $U_1$ and $U_2$ in Figures 4 and 9, respectively. $U_1$ ($U_2$) is a partition of the microdata $T_5$ in Table VIII ($T_8$ in Table XI). By Lemma 6, the probability that $Assign(T_5, 2)$ returns $U_1$ equals the probability that $Assign(T_8, 2)$ outputs $U_2$.

We prove that *Ace* ensures transparent $l$-diversity by combining Lemmas 5 and 6 with the following theorem, which states a sufficient condition for transparent $l$-diversity.

THEOREM 2. *Let $\mathcal{G}_A$ and $\mathcal{G}_B$ be two algorithms as follows:*

(1) *$\mathcal{G}_A$ takes as input a microdata table $T_1$ and a positive integer $l$, and outputs a bucket partition $U_1$ of $T_1$, such that for any bucket partition $U_2$ symmetric to $U_1$, we have $Pr\{\mathcal{G}_A(T_1, l) = U_1\} = Pr\{\mathcal{G}_A(T_2, l) = U_2\}$, where $T_2 = \bigcup_{B \in U_2} B$.*

(2) *$\mathcal{G}_B$ is a deterministic algorithm that takes as input a bucket partition $U$ and outputs another bucket partition, such that for any bucket partition $U'$ symmetric to $U$, $\mathcal{G}_B(U)$ is always symmetric to $\mathcal{G}_B(U')$.*

*Let $\mathcal{G}$ be an $l$-diversity algorithm that first applies $\mathcal{G}_A$ on the input microdata, then invokes $\mathcal{G}_B$ on the bucket partition output from $\mathcal{G}_A$, and finally returns the anonymization decided by the bucket partition generated from $\mathcal{G}_B$. $\mathcal{G}$ is transparent.*

**Algorithm** *Hybrid* $(T, l)$
1. if $T$ is not $l$-eligible then return $\emptyset$
2. $G_0$ = a QI-group containing all tuples in $T$, and $P = \{G_0\}$
3. while there exists a $2l$-diverse QI-group $G$ in $P$
4.    $\{G_a, G_b\}$ = the canonical $l$-cut of $G$
5.    $P = P - \{G\} + \{G_a, G_b\}$
6. $T^* = \emptyset$
7. for each QI-group $G_i \in P$
8.    $T_i^* = Ace(G_i, l)$
9.    $T^* = T^* \cup T_i^*$
10. return $T^*$

Fig. 10.   The *Hybrid* algorithm

By Lemma 6 (Lemma 5), *Assign* (*Slice*) satisfies the requirements for $\mathcal{G}_A$ ($\mathcal{G}_B$) stated in Theorem 2; therefore, *Ace* (as a combination of *Assign* and *Slice*) is a transparent algorithm.

### 3.3   The *Hybrid* Algorithm

This section develops a new algorithm *Hybrid* that combines *Tailor* and *Ace*. *Hybrid* is motivated by, and overcomes the drawbacks of, *Tailor* and *Ace*. We will first explain those drawbacks, and then, elaborate the details of *Hybrid*.

Given a microdata $T$ and an integer $l$, *Tailor* initiates a partition $P = \{T\}$, and then iteratively refines $P$, by splitting the QI-groups of $P$ into smaller ones. However, once a QI-group violates $2l$-diversity, it is ignored by *Tailor*, even if it can be further divided. As a result, *Tailor* sometimes spawns QI-groups with many tuples, entailing high information loss. For example, consider the 2-diverse generalization $T_6^*$ (Table IX), which is produced by *Tailor* in Example 4. The first QI-group $G_1$ in $T_6^*$ has four tuples {Ann, Bob, Cate, Don} in $T_5$ (Table VIII). In fact, $G_1$ can be further split into 2-diverse QI-groups {Ann, Cate} and {Bob, Don}. *Tailor* fails to see the split because $G_1$ is not 4-diverse.

*Ace* does not suffer from the above defect, but its random nature may occasionally create poor QI-groups. Recall that, *Ace* employs *Assign* to obtain an $l$-diverse bucket partition $U$ of $T$. Let us revisit the way *Assign* builds a bucket $B$ in $U$: *Assign* first decides the signature of $B$, and then determines each column in $B$, using tuples *randomly* selected from $T$. The distribution of QI values in each column of $B$ may vary significantly. For instance, in Example 6, *Assign* generates a bucket $B_3$ with signature {*diabetes*, *gastritis*}. Diabetes usually affects people over 40, while gastritis is common for all ages. Therefore, when *Assign* constructs the *diabetes* column, the random samples from $T$ are likely to have large *Age* values. In contrast, the *gastritis* column may contain individuals with any ages.

This (QI-distribution) difference becomes problematic in *Slice*, which *Ace* deploys to refine the bucket partition $U$ output by *Assign*. As explained in Section 3.2.1, *Slice* splits each bucket $B \in U$ into non-divisible buckets (a.k.a QI-groups), each of which has exactly one tuple from every column of $B$. If the columns of $B$ have diverse QI-distributions, the tuples in a final non-divisible QI-group may have dissimilar QI values. After anonymization, such a QI-group would incur large information loss.

| | Age | Gender | Education | Birthplace | Occupation | Income |
|---|---|---|---|---|---|---|
| Size | 79 | 2 | 17 | 57 | 50 | 50 |

Table XII.    Attribute domain sizes

*Hybrid*, as in Figure 10, remedies the deficiencies of *Tailor* and *Ace* by running the two algorithms consecutively. Specifically, *Hybrid* first computes a partition $P$ of $T$ using *Tailor*. In particular, Lines 1-5 in Figure 10 are identical to Lines 1-5 in Figure 3. As the second step, *Hybrid* treats each QI-group in $P$ as a tiny microdata table, and invokes *Ace* to generalize the QI-group (Lines 6-10).

By employing *Ace* to refine $P$, *Hybrid* outputs QI-groups with (much) fewer tuples than *Tailor*, thus avoiding the defect of *Tailor*. Meanwhile, compared to *Ace*, *Hybrid* incurs lower information loss, by executing *Ace* on each QI-group in $P$, where tuples already have similar QI values. The following theorem shows that *Hybrid* is transparent.

THEOREM 3. *Let $T$ be a microdata table, $l$ be a positive integer, and $T^*$ be any possible output of Hybrid$(T, l)$. Given any external source $E$ for $T$, we have $risk(o) \leq 1/l$ for any individual $o$.*

Finally, we point out that *Hybrid* has an $O(n^2 \log n)$ time complexity, where $n$ is the number of tuples in $T$. This follows from the $O(n^2 \log n)$ complexity of both *Tailor* and *Ace*.

## 4.  EXPERIMENTS

In the earlier sections, we have proved the privacy guarantees of our transparent algorithms. A natural question is, how do they compare with the existing solutions in terms of data utility and computation overhead (remember that no previous solution is transparent, i.e., it does not ensures anonymity, when an adversary knows the algorithm details)? In the sequel, we answer this question with empirical evidence that validates the effectiveness and efficiency of our algorithms. First, Section 4.1 clarifies the experiment settings, and then Sections 4.2 and 4.3 present detailed results.

### 4.1  Experimental Setting

Following previous work [Ghinita et al. 2007; Xiao and Tao 2007], we employ two real-world datasets, OCC and SAL, extracted from the *Integrated Public Use Microdata Series* [Ruggles et al. 2004]. Both datasets consist of 600k tuples, each containing the information of an American adult. OCC has a sensitive attribute *Occupation*, and four QI attributes, *Age*, *Gender*, *Education*, and *Birthplace*. SAL has the same QI attributes, but a different sensitive attribute *Income*. All attributes have integer domains. Table XII presents their domain sizes.

We compare our techniques (adopting the MBR function) against two *l*-diversity generalization algorithms, *Mondrian* [LeFevre et al. 2006a] and *Mask* [Wong et al. 2007]. The former is a popular technique in the literature [Byun et al. 2006; LeFevre et al. 2006b; Nergiz et al. 2007; Pei et al. 2007], due to its simplicity and effectiveness. *Mask*, on the other hand, is an existing approach that does not assume adversaries with zero algorithm knowledge (nevertheless, as explained in Sections 1 and 2.3, *Mask* is not transparent, as it can prevent only minimality attacks). We apply each algorithm to compute $l$-diverse generalizations of OCC and SAL, using various values of $l$[6]. Note that the generalizations produced by our solutions are guaranteed to be transparent $l$-diverse, whereas those by the other methods are not.

| Parameter | Values |
|---|---|
| $l$ | 6, 7, **8**, 9, 10 |
| Query dimensionality $qd$ | 2, 3, 4, **5** |
| Expected selectivity $s$ | 2%, 4%, **6%**, 8%, 10% |

Table XIII.   Parameters and Tested Values

In accordance with [Ghinita et al. 2007; Wong et al. 2007; Xiao and Tao 2007], we evaluate the utility of a generalized table $T^*$ by using it to answer count queries about the underlying microdata $T$. Each query has the form:

```
SELECT COUNT(*) FROM T
WHERE  pred(A_1^q) AND ... AND pred(A_4^q) AND pred(A^s)
```

where $pred(A)$ denotes a predicate on $A$. Predicates are generated based on two parameters: *query dimensionality $qd$* and *expected selectivity $s$*. Specifically, given $qd \in [2, 5]$ and $s \in (0, 1)$, we create a set $S_A$ that contains the sensitive attribute $A^s$ of $T$, and $qd - 1$ QI attributes randomly selected. Then, for each $A \in S_A$, we set $pred(A)$ to "$A \in I$", where $I$ is a random interval on $A$, enclosing a fraction $s^{1/qd}$ of the values in $A$. Finally, for each $A' \notin S_A$, $pred(A')$ is "$A' = *$". By requiring $qd \geq 2$ and $A^s \in S_A$, we aim to examine how well $T^*$ preserves the correlation between the QI and sensitive attributes.

On each generalized table, we process several query workloads, each of which contains 1000 queries with identical $qd$ and $s$. We gauge the utility of $T^*$ by the average workload error computed as follows. For each query, we derive its exact result $act$ from $T$, and compute an estimated answer $est$ from $T^*$ using the approximation technique in [LeFevre et al. 2006a]. The error of $est$ is defined as $\frac{|act - est|}{\max\{act, \delta\}}$, where $\delta$ is set to 0.5% of the dataset cardinality. Then, the workload error equals the average error of all queries in the workload. Note that $\delta$ is introduced to prevent the workload error from being dominated by queries with exceedingly small results (similar approaches are adopted in [Garofalakis and Kumar 2005; Vitter and Wang 1999]).

Table XIII summarizes the experiment parameters. Unless otherwise specified, we always set the parameters to their default values, i.e., the bold numbers in Table XIII. All experiments are performed on a computer with a 1.8GHz CPU and 1GB memory.

### 4.2   Utility of Generalization

The first set of experiments evaluates the information loss incurred by each algorithm. Figure 11 illustrates the results as a function of $l$. As expected, the error of all methods escalates with $l$, since a larger $l$ implies a more stringent anonymity requirement, which, in turn, demands more aggressive generalization. *Hybrid* and *Mondrian* have the best overall performance. This is a strong evidence indicating that the heuristics of *Hybrid* are highly effective. In particular, even though *Hybrid* must guarantee transparency, it still offers almost the same utility compared to *Mondrian* (which is non-transparent).

---

[6]*Mask* requires two parameters $k$ and $l$ ($k \geq l$) to generate an $l$-diverse table. We set $k = l$ in our experiments, since a smaller $k$ leads to a generalized table with higher utility, as shown in [Wong et al. 2007].

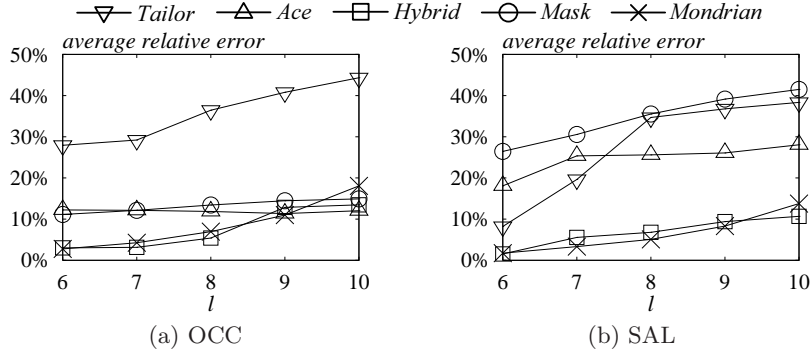

Fig. 11.   Query Accuracy vs. $l$

|             | Age  | Gender | Education | Birthplace |
|-------------|------|--------|-----------|------------|
| *Occupation* | 0.49 | 0.62   | 0.28      | 0.38       |
| *Income*     | 0.71 | 0.87   | 0.41      | 0.50       |

Table XIV.   Correlation Ratio between Attributes

*Tailor* and *Ace* exhibit worse performance than *Hybrid*. This is not surprising because, as mentioned in Section 3.3, *Hybrid* is designed to overcome the short-comings of *Tailor* and *Ace*. *Mask* incurs larger error than *Hybrid* in all cases, even though the former is vulnerable to adversaries with full algorithm knowledge (recall that *Mask* prevents only minimality attacks).

Each algorithm demonstrates similar behavior regardless of the dataset, except that *Ace* performs worse on SAL than on OCC. To explain this, we observe that the incomes depend heavily on people's ages and education. Hence, when *Ace* employs *Assign* to create a partition $U$ of SAL, each bucket in $U$ contains tuples with very different QI values, due to the reason explained in Section 3.3. As a result, the QI-groups returned by *Ace* have long generalized intervals, rendering low data utility. The above phenomenon does not exist on OCC because occupation is much less correlated to the QI-attributes. To support our analysis, Table XIV shows the correlation ratios [Kendall and Stuart 1979] between the QI and sensitive attributes of OCC and SAL. A larger ratio indicates stronger correlation.

To study the influence of query dimensionality $qd$, Figure 12 plots the workload error as a function of $qd$. The relative performance of alternative algorithms remains the same as in Figure 11. In particular, *Hybrid* and *Mondrian* permit highly accurate counting analysis; their maximum error is less than 10%. Each algorithm has better query precision when the query dimensionality $qd$ is higher. To understand this, recall that each query predicate either includes the whole domain of an attribute, or is an interval covering $s^{1/qd}$ of the domain. When $s$ is fixed but $qd$ increases, $s^{1/qd}$ becomes greater, implying wider query intervals, which lead to smaller error, as explained in [Xiao and Tao 2006a]. Figure 13 shows the error when the expected selectivity $s$ grows from 2% to 10%. Again, the relative superiority of different algorithms is the same. Their error decreases when $s$ increases, as is consistent with the experiment results in [Ghinita et al. 2007; Wong et al. 2007; Xiao and Tao 2007].

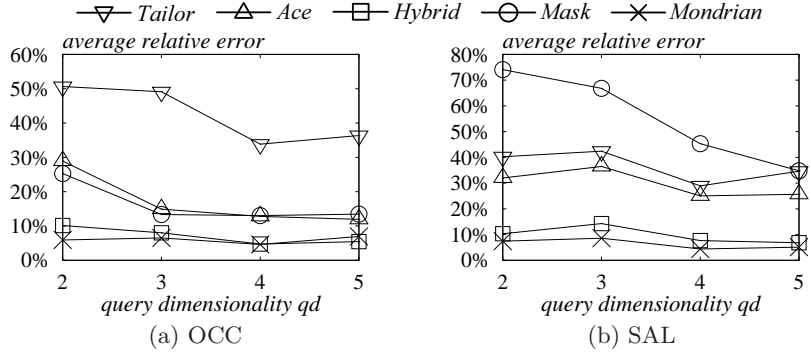

Fig. 12.   Query Accuracy vs. Query Dimensionality $qd$

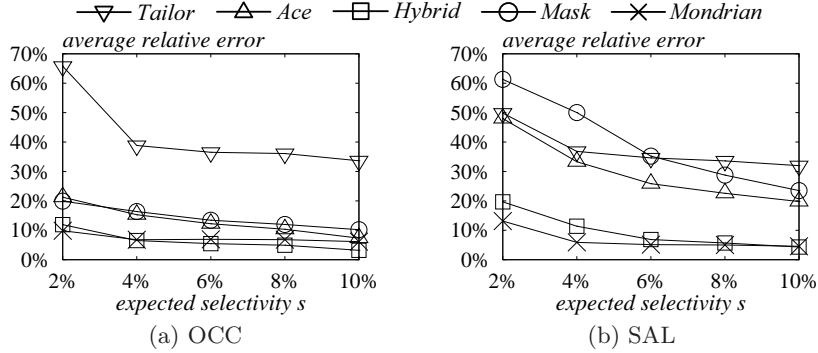

Fig. 13.   Query Accuracy vs. Expected Selectivity $s$

In summary, *Hybrid* and *Mondrian* produce generalizations with similar data utility, and both significantly outperform *Tailor*, *Ace*, and *Mask*. Therefore, overall *Hybrid* is the best anonymization technique, since it promises much stronger privacy guarantee than *Mondrian*.

## 4.3   Computation Overhead

Having examined the effectiveness of the proposed solutions, we proceed to evaluate their efficiency. In order to inspect their scalability with the dataset cardinality, based on OCC (SAL), we generate microdata tables with various cardinalities. Specifically, given a multiple $n$ of 600k, a table with $n$ tuples is synthesized by including $n/600k$ copies of OCC (SAL). Figure 14 shows the generalization time of each method, as a function of $n$. The running time of *Mask* exhibits a superlinear increase with $n$, while the other algorithms scale almost linearly. *Hybrid* requires slightly higher overhead than *Mondrian*. This is not a serious disadvantage because

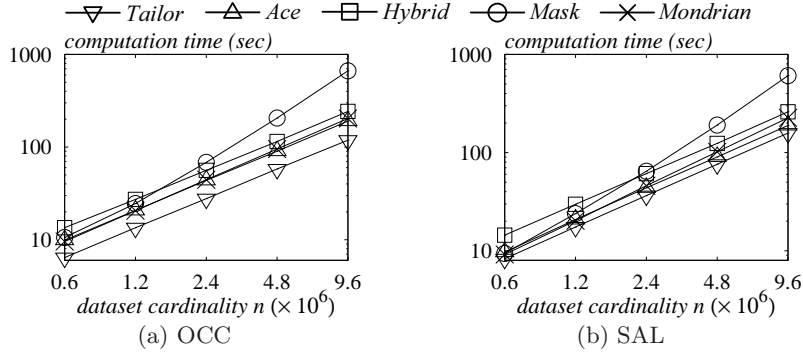

Fig. 14. Computation Time vs. Dataset Cardinality $n$

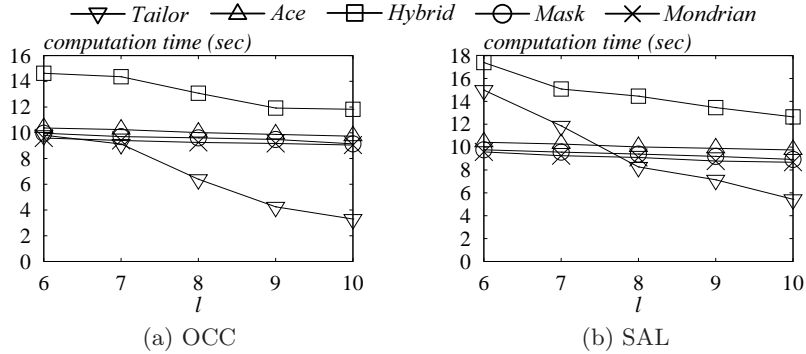

Fig. 15. Computation Time vs. $l$

(i) the difference is not large, (ii) the disadvantage is the compensated by the transparency of *Hybrid*, and (iii) anonymization is an offline process, so it is reasonable to spend a little more time preparing a publication that safeguards privacy better.

Utilizing the 600k datasets, in Figure 15, we inspect the computation overhead as a function of $l$. The running time of *Ace*, *Mondrian*, and *Mask* is insensitive to $l$. In contrast, the processing cost of *Tailor* and *Hybrid* decreases rapidly as $l$ grows. Recall that, *Tailor* works by iteratively dividing QI-groups, until all QI-groups violate $2l$-diversity. As $l$ increases, fewer $2l$-diverse QI-groups exist; hence *Tailor* terminates earlier. *Hybrid* has similar behavior because it deploys *Tailor* as the first step.

In summary, *Hybrid* is ideal for practical applications because its computation cost enjoys linear scalability to the dataset cardinality. In particular, it anonymizes a dataset with nearly 10 million tuples within 5 minutes (see Figure 14).

## 5. RELATED WORK

The works closest to ours are due to Wong et al. [2007] and Zhang et al. [2007]. Since they has been discussed extensively in Sections 1 and 2.3, the following review concentrates on the rest of the literature on privacy preserving data publishing.

A bulk of the literature focuses on designing privacy principles. The earliest principle, $k$-anonymity [Samarati 2001], requires that every QI-group should contain at least $k$ tuples. Machanavajjhala et al. [2007] point out that a $k$-anonymous table may still incur privacy breach, unless each QI-group includes sufficiently diverse sensitive values. This observation leads to the concept of $l$-diversity, which has several instantiations, e.g., entropy $l$-diversity, recursive $(c, l)$-diversity, as discussed in Section 2.1. Besides $k$-anonymity and $l$-diversity, numerous other privacy principles [Byun et al. 2006; Chen et al. 2007; Kifer and Gehrke 2006; LeFevre et al. 2006b; Li et al. 2007; Martin et al. 2007; Nergiz et al. 2007; Wang and Fung 2006; Wong et al. 2006; Xiao and Tao 2006b; 2007; Wong et al. 2007; Zhang et al. 2007; Zhang et al. 2007] have been developed to offer different flavors of privacy protection, by placing various constraints on the contents of QI-groups. Our transparent $l$-diversity principle distinguishes itself from all the previous principles, in that it guarantees privacy even when the anonymization process is public knowledge.

Generalization algorithms is another well-explored topic [Aggarwal et al. 2006; Bayardo and Agrawal 2005; Fung et al. 2005; Ghinita et al. 2007; LeFevre et al. 2005; 2006a; 2006b; Iyengar 2002; Wang et al. 2004; Wong et al. 2006; Xiao and Tao 2007; Xu et al. 2006; Wong et al. 2007; Zhang et al. 2007]. These solutions aim at minimizing the information loss, according to different anonymization constraints (e.g., global/local recoding) and measurements of loss (e.g., discernibility). Many of them are initially devised for $k$-anonymity, but can be modified to support $l$-diversity and other principles, as explained in [Machanavajjhala et al. 2007]. However, except the algorithms proposed in [Xiao and Tao 2007; Zhang et al. 2007], none of these algorithms is transparent. In other words, they can no longer ensure the privacy guarantee of the underlying principle, when an adversary is aware of the details of the algorithm.

Other problems related to generalization have also attracted considerable research efforts. Specifically, optimal $k$-anonymous generalization has been shown to be NP-hard in [Aggarwal et al. 2005; Meyerson and Williams 2004; Park and Shim 2007], which also develop approximation algorithms with provable worst-case quality guarantees. Aggarwal [2005] shows that when the number of QI attributes is large, it is simply impossible to achieve $k$-anonymity without substantial information loss (even when $k$ is small). Xiao and Tao [2006a] develop anatomy as an alternative anonymization technique that achieves higher data utility than generalization does.

In addition, there exist several anonymization techniques [Agrawal et al. 2005; Dwork et al. 2006; Evfimievski et al. 2003; Machanavajjhala et al. 2008; Tao et al. 2008] that do not adopt generalization. Instead, they anonymize microdata by adding random "noise" into the data, i.e., by replacing a fraction of tuples in the microdata with randomly generated tuples [Agrawal et al. 2005; Evfimievski et al. 2003; Tao et al. 2008], or by deriving the tuple distribution in the microdata and then publishing a noisy version of the distribution [Dwork et al. 2006; Machanavajjhala et al. 2008]. These techniques are designed by assuming that the process for generating random "noise" is known to the public, and hence, they do not suffer from reverse engineering attacks.

## 6.   CONCLUSIONS

Most existing anonymization techniques fail to protect privacy against adversaries with full knowledge of the anonymization mechanism. In this paper, we remedy the problem with two important contributions. First, we provide a thorough analysis on the disclosure risks in the anonymized tables, assuming that the anonymization algorithm is public knowledge. This analysis leads to the formulation of transparent $l$-diversity, which ensures small disclosure risks in an anonymized table, even if everything involved in the anonymization process, except the microdata, is revealed to the public. Second, we identify three anonymized algorithms that can enforce transparent $l$-diversity, and demonstrate their practical usefulness through extensive experiments.

This work also lays down a solid foundation for future research. First, our analysis focuses on $l$-diversity due to its popularity in the literature. However, the concept of transparent anonymization is general, and can be integrated with any other principle (e.g., $t$-closeness [Li et al. 2007], $\delta$-presence [Nergiz et al. 2007]). It is an interesting direction to design transparent generalization algorithms for those principles. Second, the proposed solutions are heuristic in nature, and do not have attractive asymptotical performance guarantees. It is a challenging problem to study theoretical transparent algorithms. Note that the existing findings (including the complexity results, approximation algorithms, etc.) were derived for conventional generalization, and hence, are not immediately applicable to transparent anonymization.

## APPENDIX

**Proof of Proposition 1.** Observe that, the adversary's knowledge about the external source $E$ can be expressed as $T \in S$, since $S$ consists of all microdata tables that involve the individuals in $E$. Furthermore, if $T \in S_{o,v}$, then $o$ has a sensitive value $v$ in $T$, and vice versa. Hence,

$$
\begin{aligned}
risk(o) &= \max_{v \in A^*} Pr\big\{o \text{ has } v \text{ in } T \mid E \wedge \mathcal{G} \wedge T^* \wedge l\big\} \\
&= \max_{v \in A^*} Pr\big\{T \in S_{o,v} \mid T \in S \wedge \mathcal{G} \wedge T^* \wedge l\big\} \\
&= \max_{v \in A^*} Pr\big\{T \in S_{o,v} \mid T \in S \wedge \mathcal{G}(T,l) = T^*\big\} \\
&= \max_{v \in A^*} \frac{Pr\big\{T \in S_{o,v} \wedge T \in S \wedge \mathcal{G}(T,l) = T^*\big\}}{Pr\big\{T \in S \wedge \mathcal{G}(T,l) = T^*\big\}} \\
&= \max_{v \in A^*} \frac{Pr\big\{T \in S_{o,v} \wedge \mathcal{G}(T,l) = T^*\big\}}{Pr\big\{T \in S \wedge \mathcal{G}(T,l) = T^*\big\}} \quad \text{(since } S_{o,v} \subseteq S) \\
&= \max_{v \in A^*} \frac{\sum_{\hat{T} \in S_{o,v}} \big(Pr\big\{T = \hat{T}\big\} \cdot Pr\big\{\mathcal{G}(\hat{T},l) = T^*\big\}\big)}{\sum_{\hat{T} \in S} \big(Pr\big\{T = \hat{T}\big\} \cdot Pr\big\{\mathcal{G}(\hat{T},l) = T^*\big\}\big)}.
\end{aligned}
$$

Recall that, each possible microdata instance in $S$ is equally likely for the adversary, before s/he observes $T^*$. That is, for any $\hat{T}_1, \hat{T}_2 \in S$, we have $Pr\big\{T = \hat{T}_1\big\} =$

$Pr\{T = \hat{T}_2\}$. Thus,

$$
\begin{aligned}
risk(o) &= \max_{v \in A^s} \frac{\sum_{\hat{T} \in S_{o,v}} \left( Pr\{T = \hat{T}\} \cdot Pr\{\mathcal{G}(\hat{T}, l) = T^*\} \right)}{\sum_{\hat{T} \in S} \left( Pr\{T = \hat{T}\} \cdot Pr\{\mathcal{G}(\hat{T}, l) = T^*\} \right)} \\
&= \max_{v \in A^s} \frac{\sum_{\hat{T} \in S_{o,v}} Pr\{\mathcal{G}(\hat{T}, l) = T^*\}}{\sum_{\hat{T} \in S} Pr\{\mathcal{G}(\hat{T}, l) = T^*\}},
\end{aligned}
$$

which completes the proof. □

**Proof of Lemma 1.** Assume by contradiction that *Opt-Gen* is not a minimal algorithm. Then, there exists a microdata table $T$ and a positive integer $l$, such that $T_1^* = Opt\text{-}Gen(T, l)$ is not a minimal $l$-diverse generalization of $T$, with respect to the MBR function $f$ and the global recoding scheme. Let $P_1$ be the partition of $T$ that decides $T_1^*$. By Definition 7, there should be a child $P_2$ of $P_1$, such that $P_2$ and $f$ decide a generalization $T_2^*$ that conforms to the global recoding scheme.

According to Definition 6, (i) there exists a unique QI-group $G_1$ in $P$ that does not appear in $P_2$, and (ii) $P_2$ contains only two QI-groups $G_2$ and $G_3$ that are not included in $P_1$. Furthermore, since $G_1 = G_2 \cup G_3$ and $G_2 \cap G_3 = \emptyset$, we have $|G_1| = |G_2| + |G_3|$. Thus,

$$
\begin{aligned}
\sum_{G \in P_1} |G|^2 &= |G_1|^2 + \sum_{G \in P_1 - \{G_1\}} |G|^2 \\
&\geq |G_2|^2 + |G_3|^2 + \sum_{G \in P_1 - \{G_1\}} |G|^2 \\
&= \sum_{G \in P_1 - \{G_1\} + \{G_2, G_3\}} |G|^2 \\
&= \sum_{G \in P_2} |G|^2,
\end{aligned}
$$

which contradicts the fact that *Opt-Gen* minimizes the discernability of the generalized tables. Hence, the lemma is proved. □

**Proof of Lemma 2.** Let $G_3'$ ($G_4'$) be the set of tuples in $G'$, such that $G_3'$ and $G_1$ ($G_4'$ and $G_2$) involve the same set of individuals. To prove the lemma, it suffices to show that $\{G_3', G_4'\}$ is the canonical $l$-cut of $G'$.

Without loss of generality, assume that $\{G_1, G_2\}$ is an $l$-cut of $G$ on $A_i^q$ ($i \in [1, d]$). We will first prove that $\{G_3', G_4'\}$ is an $l$-cut of $G'$ on $A_i^q$, i.e., $\{G_3', G_4'\}$ satisfies the three conditions in Definition 9. Observe that the first condition trivially holds. Let $v$ be the most frequent $A^s$ value in $G$, and $c$ be the number of tuples in $G$ with a sensitive value $v$. Since $G$ and $G'$ are isomorphic, they contain the same multi-set of $A^s$ values. Therefore, $c$ is also the maximum number of tuples in $G'$ with an identical sensitive value. Since $|G_3'| = |G_1| \geq c \cdot l$ and $|G_4'| = |G_2| \geq c \cdot l$, $\{G_3', G_4'\}$ fulfills the second condition in Definition 9.

Assume by contradiction that, $\{G_3', G_4'\}$ violates the third condition in Definition 9. There should exist $t_3' \in G_3'$ and $t_4' \in G_4'$, such that (i) $t_3'[A_i^q] > t_4'[A_i^q]$, or (ii) $t_3'[A_i^q] = t_4'[A_i^q]$ and $t_3'[A^{id}] = t_4'[A^{id}]$. Let $t_1$ ($t_2$) be the tuple in $G_1$ ($G_2$), such that $t_1$ and $t_3$ ($t_2$ and $t_4$) concern the same individual. Then, $t_1$ and $t'_3$ ($t_2$ and $t'_4$) should have the same QI values. As a result, we have either (i) $t_1[A_i^q] > t_2[A_i^q]$, or (ii) $t_1[A_i^q] = t_2[A_i^q]$ and $t_1[A^{id}] > t_2[A^{id}]$. This contradicts the assumption that $\{G_1, G_2\}$ is an $l$-cut of $G$. Therefore, $\{G'_3, G'_4\}$ is an $l$-cut of $G'$ on $A_i^q$.

Next, we will show that $\{G'_3, G'_4\}$ is canonical. Assume that this is not true. Then, by Definition 10, at least one of the following three conditions must hold:

(1) Among the $l$-cuts of $G'$, the perimeter of $\{G'_3, G'_4\}$ is not the smallest.
(2) There exists an $l$-cut $\{G'_5, G'_6\}$ of $G'$ on $A_j^q$ ($j < i$), such that $h_p(G'_5) + h_p(G'_6) = h_p(G'_3) + h_p(G'_4)$.
(3) There exists an $l$-cut $\{G'_5, G'_6\}$ of $G'$ on $A_i^q$, such that $|G'_5| < |G'_3|$, and $h_p(G'_5) + h_p(G'_6) = h_p(G'_3) + h_p(G'_4)$.

Consider that Condition 3 is satisfied. Let $G_5$ ($G_6$) be the set of tuples in $G$, such that $G_5$ and $G'_5$ ($G_6$ and $G'_6$) contain the same set of individuals. It can be verified that $\{G_5, G_6\}$ is an $l$-cut of $G$ on $A_i^q$, and $h_p(G_5) + h_p(G_6) = h_p(G'_5) + h_p(G'_6)$. Then,

$$h_p(G_5) + h_p(G_6) = h_p(G'_5) + h_p(G'_6) = h_p(G'_3) + h_p(G'_4) = h_p(G_1) + h_p(G_2).$$

Furthermore, $|G_5| = |G'_5| < |G'_3| = |G_1|$. This contradicts the assumption that $\{G_1, G_2\}$ is the canonical $l$-cut of $G$.

Similarly, it can be shown that when Condition 1 or 2 holds, $\{G_1, G_2\}$ cannot be the canonical $l$-cut of $G$, leading to a contradiction. Thus, $\{G'_3, G'_4\}$ should be the canonical $l$-cut of $G'$, which completes the proof.    □

**Proof of Lemma 3.** Let $T_2^* = \textit{Tailor}(T_2, l)$, and $P_3$ be the partition of $T_2$ that decides $T_2^*$. We will prove the lemma, by showing that (i) $P_1$ and $P_3$ are isomorphic, and (ii) $P_2 = P_3$. The former guarantees that $T_2^* = T^*$, since isomorphic partitions always lead to the same anonymization.

To facilitate our proof, we construct a binary tree $R_1$ of QI-groups as follows. First, we set the root of $R_1$ to $T_1$. Then, we apply *Tailor* on $T_1$ with the given $l$ value, and monitor the execution of *Tailor*. As shown in Figure 3, *Tailor* will first construct a partition $P = \{G_0\}$, with $G_0 = T_1$. Then, each time *Tailor* computes the canonical $l$-cut $\{G_1, G_2\}$ of QI-group $G \in P$, we insert $G_1$ and $G_2$ (into $R$) as the child nodes of $G$. As such, after *Tailor* terminates, each leaf of $R_1$ is a QI-group in $P_1$, and vice versa. We refer to $R_1$ as the *split history* of $T_1$. Following the same methodology, we also construct the split history $R_2$ of $T_2$, such that the leaves of $R_2$ constitute $P_3$.

Next, we will prove that $P_1$ is isomorphic to $P_3$, by showing that each leaf of $R_1$ is isomorphic to a leaf of $R_2$, and vice versa. Our proof is by induction. For the base case, let us consider the roots of $R_1$ and $R_2$. Let $G_1$ ($G_2$) denote the root of $R_1$ ($R_2$). We have $G_1 = T_1$ and $G_2 = T_2$. Since $P_1$ and $P_2$ are isomorphic, $T_1$ and $T_2$ should also be isomorphic, because $T_1 = \bigcup_{G \in P_1} G$ and $T_1 = \bigcup_{G \in P_2} G$. Therefore, $G_1$ is isomorphic to $G_2$.

As a second step, assume that two nodes $G_3 \in R_1$ and $G_4 \in R_2$ are isomorphic. We will establish two propositions:

—Proposition 1. $G_3$ is a leaf of $R_1$, if and only if $G_4$ is a leaf of $R_2$.

—Proposition 2. If $G_3$ is not a leaf, then each child of $G_3$ is isomorphic to a child of $G_4$.

Observe that, $G_3$ ($G_4$) is a leaf of $R_1$ ($R_2$), if and only if it is not $2l$-diverse, otherwise it would have been divided into smaller parts by *Tailor*. Since $G_3$ and $G_4$ are isomorphic, if $G_3$ is not $2l$-diverse, $G_4$ must violate $2l$-diversity, and vice versa. Therefore, Proposition 1 holds.

Now assume that $G_3$ is not a leaf. Let $\{G_a, G_b\}$ and $\{G'_a, G'_b\}$ be the canonical $l$-cuts of $G_3$ and $G_4$, respectively. By Lemma 2, $G_a$ and $G'_a$ ($G_b$ and $G'_b$) contain the same set of individuals. We will show that $G_a$ ($G_b$) is isomorphic to $G'_a$ ($G'_b$).

Consider the set $S_a$ of leaves under the subtree of $G_a$. We have $\bigcup_{G \in S_a} G = G_a$. Since $P_1$ and $P_2$ are isomorphic, there exists a subset $S'_a$ of $P_2$, such that each $G \in S_a$ is isomorphic to some $G' \in S'_a$, and vice versa. Let $G_5 = \bigcup_{G' \in S'_a} G'$. Then, $G_5$ is isomorphic to $G_a$, which indicates that $G_5$ and $G_a$ involve the same set of individuals. Recall that $G_a$ and $G'_a$ also contain an identical set of individuals. Hence, each individual in $G'_a$ appears in $G_5$, and vice versa. Because both $G'_a$ and $G_5$ are subsets of $T_2$, we have $G'_a = G_5$. Consequently, $G'_a$ is isomorphic to $G_a$. Similarly, it can be verified that the $G_b$ and $G'_b$ are isomorphic. Thus, Proposition 2 is valid. By induction, each leave of $R_1$ is isomorphic to a leaf of $R_2$, and vice versa. Hence, $P_1$ is isomorphic to $P_3$.

To complete the proof, it remains to show that $P_2 = P_3$. Since both $P_2$ and $P_3$ are isomorphic to $P_1$, $P_2$ must be isomorphic to $P_3$. Therefore, for each QI-group $G \in P_2$, there exists $G' \in P_3$, such that $G$ and $G'$ involve the same set of individuals. This indicates that $G = G'$, since the both $G$ and $G'$ are subsets of $T_2$. Therefore, $P_2 = P_3$, which proves the lemma.  $\square$

**Proof of Theorem 1.** Let $T$ be any microdata table, $l$ be any positive integer, and $T^* = \mathcal{G}(T, l)$. Let $E$ be any external source, and $C$ be the set of possible microdata instances based on $E$, such that $\mathcal{G}(\hat{T}, l) = T^*$ for any $\hat{T} \in C$. Let $o$ be any individual, $v$ be an arbitrary sensitive value, and $C'$ the subset of $C$, such that each $\hat{T} \in C'$ contains a tuple $t$ with $t[A^{id}] = o$ and $t[A^s] = v$. By Proposition 1, we can prove Theorem 1 by showing that

$$\frac{|C'|}{|C|} \leq \frac{1}{l}. \tag{5}$$

For each $\hat{T} \in C$, we define the *essential partition* of $\hat{T}$, as the partition of $\hat{T}$ generated by $\mathcal{G}$, when taking $\hat{T}$ and $l$ as input. We divide $C$ into disjoint clusters, such that each cluster is a maximal set of instances (in $C$) whose essential partitions are isomorphic. Let $n$ be the total number of clusters in $C$, and $C_j$ ($j \in [1, n]$) the $j$-th cluster. Let $C'_j$ be a set containing the instances in $C_j$ that associate $o$ with $v$. In the following, we will show that $|C'_j|/|C_j| \leq 1/l$ for any $j \in [1, n]$, which will prove the theorem, as it leads to

$$\frac{|C'|}{|C|} = \frac{\sum_{j=1}^{n} |C'_j|}{\sum_{j=1}^{n} |C_j|} \leq \frac{\sum_{j=1}^{n} |C_j|/l}{\sum_{j=1}^{n} |C_j|} = \frac{1}{l}. \tag{6}$$

Consider any $\hat{T} \in C_j$ for some $j \in [1, n]$. Let $\hat{P}$ be the essential partition of $\hat{T}$, $m = |\hat{P}|$, and $G_k$ ($k \in [1, m]$) the $k$-th QI-group in $\hat{P}$. Let $\hat{P}'$ be a partition isomor-

phic to $\hat{P}$, and $\hat{T}' = \bigcup_{G' \in \hat{P}'}$. Since $\hat{T}'$ and $\hat{T}$ involve the same set of individuals, $\hat{T}'$ is a possible microdata instance based on $E$. By the assumption on $\mathcal{G}$, we have $\mathcal{G}(\hat{T}', l) = T^*$. Therefore, $\hat{T}' \in C_j$. In other words, for any partition $\hat{P}'$ isomorphic to $\hat{P}$, the microdata corresponding to $\hat{P}'$ is contained in $C_j$. Then, by the definition of $C_j$, $|C_j|$ should equal the total number of distinct partitions isomorphic to $\hat{P}$, including $\hat{P}$ itself. According to the definition of partition isomorphism, we can obtain any partition isomorphic to $\hat{P}$, by replacing any QI-groups in $\hat{P}$ with their isomorphic counterparts. Let $a_k$ be the number of distinct QI-groups isomorphic to $G_k$. Then, the total number of partitions isomorphic to $\hat{P}$ should be $\prod_{k=1}^{m} a_k$. That is, $|C_j| = \prod_{k=1}^{m} a_k$.

Next, we will derive the value of $|C'_j|$. Without loss of generality, assume that $o$ appears in the first QI-group $G_1$ of $P_i$. Among the QI-groups isomorphic to $G_1$, let $a'_1$ be the number of QI-groups that associate $o$ with a sensitive value $v$. Then, we have $|C'_j| = a'_1 \cdot \prod_{k=2}^{m} a_k$. Therefore, $|C'_j|/|C_j| = a'_1/a_1$.

If $v$ does not appear in $G_1$, then $a'_1 = 0$. Otherwise, assume that $G_1$ contains $x$ sensitive values $v_1, v_2, ..., v_x$, such that $v_1 = v$. Further assume that, there exist $b_i$ $(i \in [1, x])$ tuples in $G_1$ with a sensitive value $v_i$. Then, there are $\frac{|G_1|!}{\sum_{i=1}^{x}(b_i!)}$ different combinations between the sensitive values and the individuals in $G_1$. Since each combination corresponds to a QI-group isomorphic to $G_1$, we have

$$a_1 = \frac{|G_1|!}{\sum_{i=1}^{x}(b_i!)}. \tag{7}$$

Observe that, among the $a_1$ combinations, there exist $\frac{(|G_1|-1)!}{\sum_{i=2}^{x}(b_i!)}$ combinations that assigns a sensitive value $v_1$ to $o$. Therefore,

$$a'_1 = \frac{(|G_1| - 1)!}{\sum_{i=2}^{x}(b_i!)}. \tag{8}$$

Hence, we have

$$\frac{|C'_j|}{|C_j|} = \frac{a'_1}{a_1} = \frac{(|G_1| - 1)! / \sum_{i=2}^{x}(b_i!)}{(|G_1|!) / \sum_{i=1}^{x}(b_i!)} = \frac{b_1}{|G_1|}. \tag{9}$$

Since $G_1$ is $l$-diverse, we have $b_1/|G_1| \le 1/l$. Consequently, $|C'_j|/|C_j| \le 1/l$, which completes the proof. □

**Proof of Lemma 4.** Given any two sets $S_t$ and $S'_t$ of tuples, we say that they are *cousins*, if $S_t$ and $S'_t$ involve the same set of individuals. To prove the lemma, we first establish the following proposition:

—Proposition 3. Let $B$ and $B'$ be two symmetric buckets, and $\{B_a, B_b\}$ be a division of $B$ on $A_i^q$ $(i \in [1, d])$. Let $B'_a$ $(B'_b)$ be the subset of $B'$, such that $B_a$ and $B'_a$ $(B_b$ and $B'_b)$ are cousins. Then, $\{B'_a, B'_b\}$ is a division of $B'$ on $A_i^q$. Furthermore, $B'_a$ and $B_a$ $(B'_b$ and $B_b)$ are symmetric.

Let $V$ be the signature of $B$, and $x = |V|$. Since $B$ and $B'$ are symmetric, $V$ should also be the signature of $B'$. Because $B'_a$ and $B'_b$ are subsets of $B'$, their signatures should be subsets of $V$. In the following, we will first show that $B'_a$ is a bucket with a signature $V$. Assume that this is not true. Then, there must exist a column $L_1'$ of $B_a'$, such that $|L_1'| \geq |B_a'|/x$. Let $L_1$ be the subset of $B_a$, such that $L_1$ and $L_1'$ are cousins. Because $B$ and $B'$ are symmetric, if any two individuals have the same sensitive value in $B'$, they should also have an identical $A^s$ value in $B$. This indicates that all tuples in $L_1$ share the same $A^s$ value. Since $|L_1| \geq |B_a|/x$, $B_a$ should have a column with more than $|B_a|/x$ tuples. This contradicts the assumption that $B_a$ is a bucket. Therefore, $B_a'$ must be a bucket with a signature $V$. By the same reasoning, it can be proved that $B_b'$ is also a bucket with a signature $V$.

Assume by contradiction that, $\{B_a', B_b'\}$ is not a division of $B'$ on $A_i^q$. Then, by Definition 11, there must exist two tuples $t_a' \in B_a'$ and $t_b' \in B_b'$, such that (i) $t_a'[A_i^q] > t_b'[A_i^q]$, or (ii) $t_a'[A_i^q] = t_b'[A_i^q]$ and $t_a'[A^{id}] > t_b'[A^{id}]$. Let $t_a$ and $t_b$ be the tuples in $B_a$, such that $t_a$ and $t_a'$ ($t_b$ and $t_b'$) concern the same individual. Then, we have either (i) $t_a[A_i^q] > t_b[A_i^q]$, or (ii) $t_a[A_i^q] = t_b[A_i^q]$ and $t_a[A^{id}] > t_b[A^{id}]$. In that case, $\{B_a, B_b\}$ is not a division of $B$, leading to a contradiction. Hence, $\{B_a', B_b'\}$ must be a division of $B'$.

To prove Proposition 3, it remains to show that $B_a'$ and $B_a$ ($B_b'$ and $B_b$) are symmetric. Consider any column $L_2 \in B_a$. We have $|L_2| = |B_a|/x = |B_a'|/x$. Let $L_2'$ be the cousin of $L_2$ in $B_a'$. Then, $|L_2'| = |L_2| = |B_a'|/x$. Since $B$ and $B'$ are symmetric, for any individuals with the same sensitive value in $B$, they should also share an identical $A^s$ value in $B'$. Therefore, all tuples in $L_2'$ have the same sensitive value. Observe that, each column in $B_a'$ should contain exactly $|B_a'|/x$ tuples, which indicates that $L_2'$ is a column in $B_a'$. In summary, for any column $L_2$ in $B_a$, there exists a column $L_2'$ in $B_a'$, such that $L_2$ and $L_2'$ involve an identical set of individuals. Hence, $B_a'$ is symmetric to $B_a$. Similarly, it can be shown that $B_b'$ and $B_b$ are symmetric. Thus, Proposition 3 holds.

Now we are ready to prove the lemma. Without loss of generality, assume that $\{B_1, B_2\}$ is a division of $B$ on $A_i^q$ ($i \in [1, d]$). Let $B_3'$ ($B_4'$) be the subset of $B'$, such that $B_3'$ and $B_1$ ($B_4'$ and $B_2$) are cousins. By Proposition 3, $\{B_3', B_4'\}$ is a division of $B'$ on $A_i^q$, and $B_3'$ ($B_4'$) is symmetric to $B_1$ ($B_2$). To establish the lemma, it suffice to show that $\{B_3', B_4'\}$ is the canonical division of $B'$. Assume, on the contrary, that $\{B_3', B_4'\}$ is not canonical. Then, by Definition 12, $\{B_3', B_4'\}$ should satisfy at least one of the following three conditions:

(1) $\{B_3', B_4'\}$ is not a division of $G'$ with the smallest perimeter.

(2) There exists a division $\{B_5', B_6'\}$ of $G'$ on $A_j^q$ ($j < i$), such that $h_p(B_5') + h_p(B_6') = h_p(B_3') + h_p(B_4')$.

(3) There exists a division $\{B_5', B_6'\}$ of $G'$ on $A_i^q$, such that $h_p(B_5') + h_p(B_6') = h_p(B_3') + h_p(B_4')$.

Assume that $\{B_3', B_4'\}$ fulfills Condition 3. Let $B_5$ ($B_6$) be subset of $B$, such that $B_5$ and $B_5'$ ($B_6$ and $B_6'$) are cousins. By Proposition 3, $\{B_5, B_6\}$ is a division of $B$ on $A_i^q$. Then, $|B_5| = |B_5'| < |B_3'| = |B_1|$. Since each individual has the same QI values in $B$ and $B'$,

$$h_p(B_5) + h_p(B_6) = h_p(B_5') + h_p(B_6') = h_p(B_3') + h_p(B_4') = h_p(B_1) + h_p(B_2).$$

In that case, $\{B_1, B_2\}$ cannot be the canonical division of $B$ (due to the existence of $\{B_5, B_6\}$), leading to a contradiction. Therefore, $\{B_3', B_4'\}$ must violate Condition

3. Similarly, it can be verified that $\{B_3', B_4'\}$ must also violate Conditions 1 and 2, i.e., $\{B_3', B_4'\}$ should be the canonical division of $B'$. Thus, the lemma is proved. □

**Proof of Lemma 5.** Consider that we apply *Slice* on $U_1$, with the given $l$ value. As shown in Figure 8, *Slice* will iteratively retrieve a bucket $B \in U_1$, compute the canonical division $\{B_a, B_b\}$ of $B$, and then replace $B$ with $B_a$ and $B_b$. This process is carried on, until the bucket partition $U_1'$ is obtained. Let $Q_1$ be the union of the canonical divisions computed by *Slice* in each iteration, and $Q_1' = Q_1 \cup U_1$. We organize the buckets in $Q_1'$ into $|U_1|$ binary trees as follows:

(1) For the $i$-th ($i \in [1, |U_1|]$) binary tree $R_i$, the root of $R_i$ is the $i$-th bucket $B_i$ in $U_1$.

(2) For any three buckets $B_1, B_2, B_3 \in Q'$, $B_2$ and $B_3$ are the child nodes of $B_1$, if and only if $\{B_2, B_3\}$ is a division of $B_1$.

We refer to $R_i$ as the *split history* of $B_i$. Notice that, $U_1'$ equals the union of the leaves of each $R_i$ ($i \in [1, |U_1|]$). Next, assume that we apply *Slice* on $U_2$. Let $B_i'$ denote the bucket in $U_2$ that is symmetric to $B_i$ ($i \in [1, |U_1|]$). Following the way $R_i$ is generated, we also construct the split history $R_i'$ of $B_i'$. Then, the leaves of all $R_i'$ ($i \in [1, |U_1|]$) constitute $U_2'$. To prove the lemma, it suffices to show that, for any $i \in [1, |U_1|]$, each leaf of $R_i$ is symmetric to a leaf of $R_i'$, and vice versa.

Our proof is by induction. For the base case, the root $B_i$ of $R_i$ is symmetric to the root $B_i'$ of $R_i'$. Next, assume that two nodes $B \in R_i$ and $B' \in R_i'$ are symmetric. We will show that (i) $B$ is a leaf of $R_i$, if and only if $B'$ is a leaf of $R_i'$; (ii) if $B$ is not a leaf, then each child node of $B$ is symmetric to a child node of $B'$.

As shown in Figure 8, a bucket in $R_i$ or $R_i'$ is a leaf, if and only if it is not divisible, otherwise it would have been split into two smaller buckets by *Slice*. Because $B$ and $B'$ are symmetric, all columns in $B$ and $B'$ have an equal size. Thus, $B$ is not divisible, if and only if $B'$ is not divisible. Hence, $B$ is a leaf, if and only if $B'$ is a leaf.

Next, consider that $B$ is not a leaf. Let $\{B_a, B_b\}$ and $\{B_a', B_b'\}$ be the canonical divisions of $B$ and $B'$, respectively. By Lemma 4, $B_a$ and $B_a'$ ($B_b$ and $B_b'$) must be symmetric. Therefore, each child node of $B$ is symmetric to a child node of $B'$. By induction, it can be shown that each leaf of $R_i$ is symmetric to a leaf of $R_i'$, and vice versa. Hence, the lemma is proved. □

**Proof of Lemma 6.** Consider that we apply *Assign* on $T_1$ with the given $l$ value. As shown in Figure 6, *Assign* first initializes a set $S_t = T_1$, and then iteratively creates buckets using tuples in $S_t$. Let $U$ be the partition returned by *Assign* at the end, $B_i$ the bucket constructed in the $i$-th iteration, and $S_i$ the set of tuples in $S_t$ right before the $i$-th iteration. Next, assume that we run *Assign* on $T_2$. Let $U'$ be the partition of $T_2$ generated by *Assign*, $B_i'$ the bucket created in the $i$-th iteration, and $S_i'$ the set of tuples in $S_t$ prior to the $i$-th iteration. For simplicity, we say that two buckets are *siblings*, if and only if they have the same size and the same signature. In the following, we will first prove a proposition:

—Proposition 4. for any $i \in [1, |U|]$, $B_i$ and $B_i'$ are siblings.

Consider that $i = 1$. By Lines 4-12 in Figure 6, the signature of $B_1$ should contain the $\beta$ most frequent sensitive values in $S_t$, and $|B_1| = \alpha \cdot \beta$, where the values of $\alpha$ and $\beta$ are decided by $|S_1|$ and the frequencies of sensitive values in $S_1$. The above statement still holds, if we change $B_1$ to $B'_1$, and $S_1$ to $S'_1$. Recall that $S_1 = T_1 = \bigcup_{B \in U_1} B$ and $S'_1 = T_2 = \bigcup_{B' \in U_2} B'$. Since $U_1$ and $U_2$ are symmetric, $S_1$ and $S'_1$ should have the same size, and include an identical multi-set of sensitive values. Therefore, $Assign$ should employ the same $\alpha$ and $\beta$ values to construct $B_1$ and $B'_1$. Thus, $B_1$ and $B'_1$ are siblings. Furthermore, because $S_2 = S_1 - B_1$ and $S'_2 = S'_1 - B'_1$, $S_2$ and $S'_2$ should have an equal size, and contain the same multi-set of sensitive values. In turn, this indicates that, $Assign$ should use identical $\alpha$ and $\beta$ values to generate $B_2$ and $B'_2$, i.e., $B_2$ and $B'_2$ are also siblings. By an induction on $i$, it can be shown that Proposition 4 holds.

To prove the lemma, we regard $U$ and $U'$ as random variables, and show that $Pr\{U = U_1\} = Pr\{U' = U_2\}$. Let us derive $Pr\{U = U_1\}$ first. Recall that each bucket in $U$ is constructed using tuples randomly selected from $T_1$. Therefore, $Pr\{U = U_1\}$ should equal $1/m$, where $m$ is the total number of possible ways to assign the tuples in $T_1$ into the buckets in $U$. Assume that $T_1$ contains $w$ sensitive values $v_1$, $v_2$, ..., $v_w$. Let $n_j$ ($j \in [1, w]$) be the frequency of $v_j$ in in $T_1$. Let $d_{ij}$ denote the number of tuples in $B_i$ with sensitive value $v_j$. For simplicity, define $0! = 1$. We have

$$m = \frac{\prod_{j=1}^{w}(n_j!)}{\prod_{i=1}^{|U|}\prod_{j=1}^{w}(d_{ij}!)}. \qquad (10)$$

Next, we will calculate $Pr\{U' = U_2\}$. Since $T_1$ and $T_2$ contain the same multi-set of sensitive values, for any $j \in [1, w]$, the frequency of $v_j$ in $T_2$ is also $n_j$. Furthermore, because $B_i$ and $B'_i$ ($i \in [1, |U|]$) are siblings, there should exist $d_{ij}$ tuples in $B'_i$ that have a sensitive value $v_j$. As a result, there are also $m$ distinct ways to assign the tuples in $T_2$ to the buckets in $U'$. Therefore, $Pr\{U' = U_2\} = 1/m = Pr\{U = U_1\}$, which completes the proof. □

**Proof of Theorem 2.** Let $T$ be any microdata, $l$ be any positive integer, and $T^*$ be a possible output of $\mathcal{G}$. Let $E$ be any external source, and $S$ be the set of possible microdata instances based on $E$. Let $o$ be any individual, $v$ be an arbitrary sensitive value, and $S_{o,v}$ be the subset of $S$, such that each $\hat{T} \in S_{o,v}$ associates $o$ with $v$. According to Proposition 1, we can prove Theorem 2 by showing that

$$\frac{\sum_{\hat{T} \in S_{o,v}} Pr\{\mathcal{G}(\hat{T}, l) = T^*\}}{\sum_{\hat{T} \in S} Pr\{\mathcal{G}(\hat{T}, l) = T^*\}} \leq \frac{1}{l}. \qquad (11)$$

We say that a bucket partition $U$ is a *valid partition*, if $T^*$ can be decided by the partition $U' = \mathcal{G}_B(U)$. Let $M$ be the set of all valid partitions, such that for each $U \in M$, we have $Pr\{\mathcal{G}_A(\hat{T}, l) = U\} > 0$ for some $\hat{T} \in S$. Then,

$$\frac{\sum_{\hat{T} \in S_{o,v}} Pr\{\mathcal{G}(\hat{T}, l) = T^*\}}{\sum_{\hat{T} \in S} Pr\{\mathcal{G}(\hat{T}, l) = T^*\}} = \frac{\sum_{\hat{T} \in S_{o,v}}\sum_{U \in M} Pr\{\mathcal{G}_A(\hat{T}, l) = U\}}{\sum_{\hat{T} \in S}\sum_{U \in M} Pr\{\mathcal{G}_A(\hat{T}, l) = U\}}. \qquad (12)$$

We define a bucket partition $U \in M$ as a *breaching partition*, if any QI-group $G \in U$ contains a tuple $t$, such that $t[A^{id}] = o$ and $t[A^s] = v$. Observe that, for any $\hat{T} \in S_{o,v}$, if $U$ is not a breaching partition, then $Pr\{\mathcal{G}_A(\hat{T}, l) = U\} = 0$. We divide $M$ into disjoint clusters, such that each cluster is a maximal subset of symmetric bucket partitions in $M$. Let $n$ be the total number of clusters in $M$, and $M_j$ ($j \in [1, n]$) be the $j$-th cluster. Let $M_j'$ be the set of breaching partitions in $M_j$. We have

$$\frac{\sum_{\hat{T} \in S_{o,v}} Pr\{\mathcal{G}(\hat{T}, l) = T^*\}}{\sum_{\hat{T} \in S} Pr\{\mathcal{G}(\hat{T}, l) = T^*\}} = \frac{\sum_{j=1}^{n} \sum_{U \in M_j'} \sum_{\hat{T} \in S_{o,v}} Pr\{\mathcal{G}_A(\hat{T}, l) = U\}}{\sum_{j=1}^{n} \sum_{U \in M_j} \sum_{\hat{T} \in S} Pr\{\mathcal{G}_A(\hat{T}, l) = U\}}. \quad (13)$$

For simplicity, let $p(U, \hat{T})$ denote $Pr\{\mathcal{G}_A(\hat{T}, l) = U\}$, and $q(M, S)$ denote $\sum_{U \in M} \sum_{\hat{T} \in S} p(U, \hat{T})$. We will show that $q(M_j', S_{o,v})/q(M_j, S) \leq 1/l$ for any $j \in [1, n]$. This will lead to

$$\begin{aligned}
\frac{\sum_{\hat{T} \in S_{o,v}} Pr\{\mathcal{G}(\hat{T}, l) = T^*\}}{\sum_{\hat{T} \in S} Pr\{\mathcal{G}(\hat{T}, l) = T^*\}} &= \frac{\sum_{j=1}^{n} q(M_j', S_{o,v})}{\sum_{j=1}^{n} q(M_j, S)} \\
&\leq \frac{\sum_{j=1}^{n} q(M_j, S)/l}{\sum_{j=1}^{n} q(M_j, S)} = \frac{1}{l},
\end{aligned} \quad (14)$$

which proves the theorem.

Without loss of generality, consider that $j = 1$. Let $U_k$ be the $k$-th ($k \in [1, |M_1|]$) partition in $M_1$, and $T_k = \bigcup_{B \in U_k} B$. For any microdata $\hat{T}$ different from $T_k$, we have $p(U_k, \hat{T}) = 0$, since $U_k$ is not a partition of $\hat{T}$. Therefore, $T_k \in S$ should hold, otherwise $p(U_k, \hat{T}) = 0$ for all $\hat{T} \in S$, which contradicts the assumption that $U_k \in M$. Thus, $\sum_{\hat{T} \in S} p(U_k, \hat{T}) = p(U_k, T_k)$. By our assumption on $\mathcal{G}_A$, for any $k_1, k_2 \in [1, |M_1|]$, we have $p(U_{k_1}, T_{k_1}) = p(U_{k_2}, T_{k_2})$. Hence,

$$q(M_1, S) = \sum_{j=1}^{|M_1|} \sum_{\hat{T} \in S} p(U_j, \hat{T}) = \sum_{j=1}^{|M_1|} p(U_j, T_j) = |M_1| \cdot p(U_k, T_k).$$

Similarly, it can be verified that $q(M_1', S_{o,v}) = |M_1'| \cdot p(U_k, T_k)$. Therefore, $q(M_j', S_{o,v})/q(M_j, S) = |M_1'|/|M_1|$.

Next, we will derive the value of $|M_1|$. Let $U_s$ be any partition symmetric to $U_k$, and $T_s = \bigcup_{B \in U_s} B$. Then, $T_s$ and $T_k$ should contain the same set of individuals. Hence, $T_s \in S$. Since $U_s$ and $U_k$ are symmetric, $p(U_s, T_s) = p(U_k, T_k) > 0$ holds. Therefore, $U_s$ is a valid partition. Let $U_s' = \mathcal{G}_B(U_s)$, and $U_k' = \mathcal{G}_B(U_k)$. By our assumption on $\mathcal{G}_B$, $U_s'$ and $U_k'$ are symmetric. Observe that symmetric partitions are isomorphic, and thus, they always lead to the same anonymization. Since $U_k'$ and $f$ decides $T^*$, $U_s'$ and $f$ should also determine $T^*$, which indicates that $U_s \in M$. In other words, any bucket symmetric to $U_k$ should be contained in $M$. Consequently, by the definition of $M_1$, $|M_1|$ equals the total number of partitions symmetric to $U_k$.

By Definition 13, we can obtain any partition symmetric to $U_k$, by substituting any buckets in $U_k$ with their symmetric counterparts. Let $B_i$ be the $i$-th ($i \in [1, |U_k|]$) bucket in $U_k$, and $\alpha_i$ be the number of buckets symmetric to $B_i$. Then, $|M_1| = \prod_{i=1}^{|U_k|} \alpha_i$. Without loss of generality, assume that $o$ appears in $B_1$. Among all buckets symmetric to $B_1$, let $\alpha_1'$ be the number of them that contain a tuple $t$, with $t[A^{id}] = o$ and $t[A^s] = v$. We have $|M_1| = \alpha_1' \cdot \prod_{i=2}^{|U_k|} \alpha_i$. Therefore, $|M_1'|/|M_1| = \alpha_1'/\alpha_1$.

Assume $B_1$ has a signature $V$ with $x$ sensitive values. If $v \notin V$, then $\alpha_1' = 0$. Consider that $v \in V$. Recall that, we can transform $B_1$ into any bucket symmetric to $B_1$, by swapping the sensitive values between different columns of $B_1$. Totally, there are $x!$ distinct ways to assign $x$ sensitive values to the $x$ columns of $B_1$. Because each of these assignment corresponds to bucket symmetric to $B_1$, we have $\alpha_1 = x!$. Next, consider that we assign an $A^s$ value $v$ to the column that $o$ appears. The other $x - 1$ sensitive values can be assigned in $(x - 1)!$ different manner, i.e., $\alpha_1' = (x-1)!$. Hence, $\alpha_1'/\alpha_1 = 1/x$. According to the way *Assign* constructs each bucket, we have $x \geq l$. Therefore, $|M_1'|/|M_1| = \alpha_1'/\alpha_1 \leq 1/l$, which completes the proof. $\qquad\square$

**Proof of Theorem 3.** Let $S$ the set of possible microdata instances based on $E$, and $v$ be an arbitrary sensitive value. Let $S_{o,v}$ be the subset of $S$, such that each $\hat{T} \in S_{o,v}$ involves $o$, and sets $v$ as the $A^s$ value of $o$. By Proposition 1, Theorem 3 holds if and only if

$$\frac{\sum_{\hat{T} \in S_{o,v}} Pr\{Hybrid(\hat{T}, l) = T^*\}}{\sum_{\hat{T} \in S} Pr\{Hybrid(\hat{T}, l) = T^*\}} \leq \frac{1}{l}. \tag{15}$$

Consider that we apply *Hybrid* on any $\hat{T} \in S$, with the given $l$ value. *Hybrid* first employs *Tailor* to obtain a partition $P$ of $\hat{T}$. We define $P$ as the *essential partition* of $\hat{T}$, and use $G_j$ to denote the $j$-th ($j \in [1, |P|]$) QI-group in $P$. Then, *Hybrid* invokes *Ace* to transform each $G_j \in P$ into a set $T_j^*$ of anonymized tuples. We define the ordered set $\{T_1^*, T_2^*, ..., T_{|P|}^*\}$ as a *decomposition* of $P$. Since *Ace* is a randomized algorithm, there may exist multiple decompositions of $P$. At last, *Hybrid* returns the union $T^*$ of all $T_j^*$. We use $\gamma(P, T^*)$ to denote the probability that *Hybrid* transforms $P$ into $T^*$.

Let $Q$ ($Q'$) be a set that includes the essential partition of any $\hat{T} \in S$ ($\hat{T} \in S_{o,v}$). We divide $Q$ into several clusters, such that each cluster is a maximal set of isomorphic partitions in $Q$. Let $n$ be the total number of clusters in $Q$, and $C_k$ ($k \in [1, n]$) be the $k$-th cluster. Let $C_k' = C_k \cap Q'$. Then, we have

$$\frac{\sum_{\hat{T} \in S_{o,v}} Pr\{Hybrid(\hat{T}, l) = T^*\}}{\sum_{\hat{T} \in S} Pr\{Hybrid(\hat{T}, l) = T^*\}} = \frac{\sum_{j=1}^{n} \sum_{P \in C_k'} \gamma(P, T^*)}{\sum_{k=1}^{n} \sum_{P \in C_k} \gamma(P, T^*)}. \tag{16}$$

We will prove that, for any $k \in [1, n]$,

$$\frac{\sum_{P \in C_k'} \gamma(P, T^*)}{\sum_{P \in C_k} \gamma(P, T^*)} \leq \frac{1}{l}. \tag{17}$$

This will establish the Theorem, since it ensures that

$$\begin{aligned}
\frac{\sum_{\hat{T} \in S_{o,v}} Pr\{Hybrid(\hat{T}, l) = T^*\}}{\sum_{\hat{T} \in S} Pr\{Hybrid(\hat{T}, l) = T^*\}} &= \frac{\sum_{j=1}^{n} \sum_{P \in C_k'} \gamma(P, T^*)}{\sum_{k=1}^{n} \sum_{P \in C_k} \gamma(P, T^*)} \\
&\leq \frac{\sum_{j=1}^{n} \sum_{P \in C_k} \gamma(P, T^*)/l}{\sum_{k=1}^{n} \sum_{P \in C_k} \gamma(P, T^*)} = \frac{1}{l}.
\end{aligned}$$

Without loss of generality, consider that $k = 1$. Let $P$ be an arbitrary partition in $C_1$, and $G_j$ the $j$-th QI-group in $P$. Assume that $o$ is involved in $G_1$. Further assume that, for any $P' \in C_1$, the $j$-th $(j \in [1, |P|])$ QI-group in $P'$ is isomorphic to $G_j$. Then, for any $P' \in C_1$, $o$ should appear in the first QI-group of $P'$. We split $C_1$ into sub-clusters, such that for any two partitions in the same sub-cluster, they coincide on all but the first QI-group. Let $n'$ be the number of sub-clusters in $C_1$, $D_i$ $(i \in [1, n'])$ be the $i$-th sub-cluster, and $D_i' = D_i \cap Q'$. To prove that Equation 17 holds for $k = 1$, it suffices to show that, for any $i \in [1, n']$,

$$\frac{\sum_{P \in D_i'} \gamma(P, T^*)}{\sum_{P \in D_i} \gamma(P, T^*)} \leq \frac{1}{l}. \tag{18}$$

This is because, once the above inequality is established, we have

$$
\begin{aligned}
\frac{\sum_{P \in C_1'} \gamma(P, T^*)}{\sum_{P \in C_1} \gamma(P, T^*)} &= \frac{\sum_{i=1}^{n'} \sum_{P \in D_i'} \gamma(P, T^*)}{\sum_{i=1}^{n'} \sum_{P \in D_i} \gamma(P, T^*)} \\
&\leq \frac{\sum_{i=1}^{n'} \sum_{P \in D_i} \gamma(P, T^*)/l}{\sum_{i=1}^{n'} \sum_{P \in D_i} \gamma(P, T^*)} = \frac{1}{l}.
\end{aligned} \tag{19}
$$

Assume, without loss of generality, that $i = 1$. Let $P_x$ $(x \in [1, |D_1|])$ be $x$-th partition in $D_1$, and $G_{xm}$ be the $m$-th $(m \in [1, |P_x|])$ QI-group in $P_x$. Let $S_d$ be a set containing any decomposition of any $P_x \in D_1$, and $\Omega$ be the subset of $S_d$, such that each decomposition $W \in \Omega$ leads to $T^*$, i.e., $\bigcup_{T_s^* \in W} T_s^* = T^*$. Let $W_j$ be the $j$-th decomposition in $\Omega$, and $T_{jm}^*$ the $m$-th set of anonymized tuples in $W_j$. Observe that $|W_j| = |P_x|$ for any $j \in [1, |\Omega|]$ and any $x \in [1, |D_1|]$. By the definition of $\gamma(P_x, T^*)$, we have

$$\gamma(P_x, T^*) = \sum_{j=1}^{|\Omega|} \prod_{m=1}^{|P_x|} Pr\{Ace(G_{xm}, l) = T_{jm}^*\} \tag{20}$$

For simplicity, we denote $\prod_{m=1}^{|P_x|} Pr\{Ace(G_{xm}, l) = T_{jm}^*\}$ as $p(P_x, W_j)$. Then,

$$\frac{\sum_{P \in D_1'} \gamma(P, T^*)}{\sum_{P \in D_1} \gamma(P, T^*)} = \frac{\sum_{P_x \in D_1'} \sum_{j=1}^{|\Omega|} p(P_x, W_j)}{\sum_{P_x \in D_1} \sum_{j=1}^{|\Omega|} p(P_x, W_j)}. \tag{21}$$

To prove that Equation 18 is valid when $i = 1$, we will show that

$$\frac{\sum_{P_x \in D_1'} p(P_x, W_j)}{\sum_{P_x \in D_1} p(P_x, W_j)} \leq \frac{1}{l}, \tag{22}$$

for any $j \in [1, |\Omega|]$. In particular, the above inequality ensures that

$$
\begin{aligned}
\frac{\sum_{P \in D_i'} \gamma(P, T^*)}{\sum_{P \in D_i} \gamma(P, T^*)} &= \frac{\sum_{P_x \in D_1'} \sum_{j=1}^{|\Omega|} p(P_x, W_j)}{\sum_{P_x \in D_1} \sum_{j=1}^{|\Omega|} p(P_x, W_j)} \\
&\qquad \text{(by Equation 21)} \\
&\leq \frac{\sum_{j=1}^{|\Omega|} \sum_{P_x \in D_1} p(P_x, W_j)/l}{\sum_{j=1}^{|\Omega|} \sum_{P_x \in D_1} p(P_x, W_j)} = \frac{1}{l}
\end{aligned} \tag{23}
$$

Let $q(G_{xm}, T_{jm}^*)$ denote $Pr\{Ace(G_{xm}, l) = T_{jm}^*\}$. Recall that any two partitions in $D_1$ coincide on all but the first QI-group. Therefore, given any $m \in [2, |P_x|]$ and any $j \in [1, |\Omega|]$, the value of $q(G_{xm}, T_{jm}^*)$ is fixed for all $P_x \in D_1$. Let $r_j$ denote $\prod_{k=2}^{|P_x|} q(G_{xm}, T_{jm}^*)$. Then,

$$
\begin{aligned}
p(P_x, W_j) &= \prod_{m=1}^{|P_x|} Pr\{Ace(G_{xm}, l) = T_{jm}^*\} \\
&= \prod_{m=1}^{|P_x|} q(G_{xm}, T_{jm}^*) \\
&= r_j \cdot q(G_{x1}, T_{j1}^*).
\end{aligned}
\tag{24}
$$

Therefore, for any $j \in [1, |\Omega|]$,

$$
\begin{aligned}
\frac{\sum_{P_x \in D_1'} p(P_x, W_j)}{\sum_{P_x \in D_1} p(P_x, W_j)} &= \frac{\sum_{P_x \in D_1'} \left(r_j \cdot q(G_{x1}, T_{j1}^*)\right)}{\sum_{P \in D_1} \left(r_j \cdot q(G_{x1}, T_{j1}^*)\right)} \\
&= \frac{\sum_{P_x \in D_1'} q(G_{x1}, T_{j1}^*)}{\sum_{P_x \in D_1} q(G_{x1}, T_{j1}^*)}.
\end{aligned}
\tag{25}
$$

Consequently, to prove Equation 22, it suffices to show that

$$
\frac{\sum_{P_x \in D_1'} q(G_{x1}, T_{j1}^*)}{\sum_{P_x \in D_1} q(G_{x1}, T_{j1}^*)} \leq \frac{1}{l},
\tag{26}
$$

for any $j \in [1, |\Omega|]$.

Let $S_1$ be a set containing all $G_{x1}$ ($x \in [1, |D_1|]$), and $S_1'$ the maximal subset of $S_1$, such that each $G \in S_1'$ contains a tuple $t$ with $t[A^{id}] = o$ and $t[A^s] = v$. Then,

$$
\frac{\sum_{P_x \in D_1'} q(G_{x1}, T_{j1}^*)}{\sum_{P_x \in D_1} q(G_{x1}, T_{j1}^*)} = \frac{\sum_{G \in S_1'} q(G, T_{j1}^*)}{\sum_{G \in S_1} q(G, T_{j1}^*)}.
\tag{27}
$$

By the definition of $D_1$, all QI-groups in $S_1$ are isomorphic. Therefore, all QI-groups in $S_1$ have the same projection on the identifier and QI attributes. Denote this projection as $E$. If we regard each QI-group $G_{x1} \in S_1$ as a tiny microdata table, then $E$ can be deemed as an external source for $G_{x1}$. Let $S_2$ be the set of all possible instances based on $E$. Let $S_2'$ be the set of instances in $S_2$ that contain a tuple $t$, with $t[A^{id}] = o$ and $t[A^s] = v$. By Theorem 2, given $E$ as the external source, any $T_{j1}^*$ ($j \in [1, |\Omega|]$) ensures that the disclosure risk of $o$ is at most $1/l$, i.e.,

$$
\frac{\sum_{G \in S_2'} q(G, T_{j1}^*)}{\sum_{G \in S_2} q(G, T_{j1}^*)} \leq \frac{1}{l}.
\tag{28}
$$

By Equations 27 and 28, we can establish Equation 26 by showing that

$$
\frac{\sum_{G \in S_1'} q(G, T_{j1}^*)}{\sum_{G \in S_1} q(G, T_{j1}^*)} = \frac{\sum_{G \in S_2'} q(G, T_{j1}^*)}{\sum_{G \in S_2} q(G, T_{j1}^*)}.
\tag{29}
$$

For this purpose, it suffices to prove that $q(G, T_{j1}^*) = 0$, for any $G \in (S_1 - S_2) \cup (S_2 - S_1)$ and any $G \in (S_1' - S_2') \cup (S_2' - S_1')$.

Since $S_2$ contains all microdata instances based on $E$, we have $G_{x1} \in S_1$ for any $x \in [1, |D_1|]$. Therefore, $S_1 \subseteq S_2$, which indicates that $S_1' \subseteq S_2'$. Hence, $S_1 - S_2 = S_1' - S_2' = \emptyset$. Now consider any $P_x \in D_1$ $(x \in [1, |D_1|])$. Assume that we construct a partition $P_x'$ from $P_x$, by replacing $G_{x1}$ with any of its isomorphic counterparts. Then, $P_x'$ should be isomorphic to $P_x$. By Lemma 3, $P_x'$ is an essential partition of some $\hat{T} \in S$, i.e., $P_x' \in Q$. Since $P_x'$ and $P_x$ are isomorphic, and coincide on all but the first QI-group, $P_x' \in D_1$ holds. In other words, for any $G$ isomorphic to $G_{x1}$, there exists a partition $P_i \in D_1$ $(i \in [1, |D_1|])$, such that $G = G_{i1}$. Hence, $S_1$ contains any QI-group isomorphic to $G_{x1}$.

Recall that, any $T_{j1}^*$ $(j \in [1, |\Omega|])$ is a anonymization of a certain QI-group in $S_1$. Since all QI-groups in $S_1$ are isomorphic, they contain the same multi-set of sensitive values. This indicates that any $T_{j1}^*$ and any $G_{x1}$ $(x \in [1, |D_1|])$ have an identical multi-set of sensitive values. Let $G'$ be a QI-group, such that $G' \in S_2 - S_1$. Then, $G'$ and $G_{x1}$ are not isomorphic, but involve the same set of individuals. Therefore, $G'$ and $G_{x1}$ must contain distinct multi-sets of $A^s$ values. Hence, for any $j \in [1, |\Omega|]$, the multi-sets of sensitive values in $G'$ and $T_j^*$ are different, i.e., $G'$ cannot be anonymized to $T_j^*$. Therefore, $q(G, T_{j1}^*) = 0$, for any $G \in S_1 - S_2$. Similarly, it can be shown that $q(G, T_1^*) = 0$, for any $G \in S_2' - S_1'$. Thus, Equation 29 is valid. In turn, this establishes Equations 29, 26, 22, 18, and 17. Hence, the theorem is proved. □

## ELECTRONIC APPENDIX

The electronic appendix for this article can be accessed in the ACM Digital Library by visiting the following URL: `http://www.acm.org/pubs/citations/journals/tods/20YY-V-N/p1-`.

# Transparent Anonymization:    Thwarting Adversaries Who Know the Algorithm

XIAOKUI XIAO
Nanyang Technological University
YUFEI TAO
Chinese University of Hong Kong
and
NICK KOUDAS
University of Toronto

In this electronic appendix, we exemplify an attack against the *Mask* algorithm [Wong et al. 2007], which is designed under the credibility model [Wong et al. 2007]. Figure 1 illustrates the pseudo-code of *Mask*. The algorithm takes as input a microdata table $T$, two positive integers $k$ and $l$, and a subset $V$ of the $A^s$ values. It aims to ensure that, for any individual $o$ and any sensitive value $v \in V$, the adversary would have at most $1/l$ posterior belief in the event that "$o$ appears in $T$ and has a sensitive value $v$". We will explain the details of *Mask* using an example.

EXAMPLE 1. Suppose that we apply *Mask* on the microdata $T_9$ in Table I, by setting $k = l = 2$ and $V = \{dyspepsia\}$. *Mask* first generates a $k$-anonymous partition $P$ of $T_9$, using any of the existing $k$-anonymity algorithms (Line 1 in Figure 1). Assume that $P$ contains three QI-groups, namely, {Ann, Bob}, {Cate, Don}, and {Ed, Fred, Gill}. Next, *Mask* divides $P$ into two disjoint subsets $P_1$ and $P_2$ (Lines 2-6). In particular, $P_1$ contains all the QI-groups $G$ in $P$, such that at least one sensitive value in $V$ appears more than $|G|/l$ times in $G$. Meanwhile, $P_2 = P - P_1$. In our example, $P_1$ contains only one QI-group $G' = \{$Ann, Bob$\}$.

After that, *Mask* randomly chooses a QI-group $G^+$ from $P_2$, and then modifies the sensitive values in $G'$, so that $G'$ and $G^+$ have the same sensitive value distribution (Lines 7-9). Assume that $G^+ = \{$Cate, Don$\}$. Then, $G'$ will be modified in a way, such that 50% of tuples in $G'$ would have a sensitive value *flu*, and the other 50% would have *dyspepsia*. Table II illustrates a possible result of the modification. Finally, *Mask* returns the anonymization decided by the modified partition and an anonymization function (say, the MBR function), as illustrated in Table III.    □

As $T_{10}^*$ (in Table III) is produced by *Mask* with $l = 2$, under the credibility model, an adversary has at most $1/2$ posterior belief in the event that "Ann has dyspepsia

**Algorithm** *Mask* $(T, k, l, V)$
1.  generate a $k$-anonymous partition $P$ of $T$
2.  $P_1 = P_2 = \emptyset$
3.  for each QI-group $G \in P$
4.      if one of the sensitive value in $V$ appears more than $|G|/l$ times in $G$
5.          insert $G$ into $P_1$
6.      else insert $G$ into $P_2$
7.  for each QI-group $G' \in P_1$
8.      randomly choose a QI-group $G^+ \in P_2$
9.      modify the sensitive values in $G'$, such that the distribution of sensitive values in
            $G'$ becomes the same as that in $G^+$
10. return the anonymization decided by $P_1 \cup P_2$ and an anonymization function

Fig. 1.   The *Mask* algorithm

| Name | Age | Disease |
|------|-----|---------|
| Ann | 21 | dyspepsia |
| Bob | 27 | dyspepsia |
| Cate | 32 | dyspepsia |
| Don | 32 | flu |
| Ed | 54 | flu |
| Fred | 60 | flu |
| Gill | 60 | flu |

Table I.   Microdata $T_9$

| Name | Age | Disease |
|------|-----|---------|
| Ann | 21 | flu |
| Bob | 27 | dyspepsia |
| Cate | 32 | dyspepsia |
| Don | 32 | flu |
| Ed | 54 | flu |
| Fred | 60 | flu |
| Gill | 60 | flu |

Table II.   Partition $P'$

| Age | Disease |
|-----|---------|
| [21, 27] | flu |
| [21, 27] | dyspepsia |
| 32 | dyspepsia |
| 32 | flu |
| [54, 60] | flu |
| [54, 60] | flu |
| [54, 60] | flu |

Table III.   Generalization $T_{10}^*$

in the microdata". In the following, however, we will show that the posterior belief of the adversary can be boosted to 5/8, if s/he has (i) the details of *Mask*, (ii) the parameters $k$, $l$, and $V$ with which $T_{10}^*$ is computed, and (iii) an external source that contains only the seven individuals in $T_9$.

Upon observing $T_{10}^*$, the adversary knows that $T_{10}^*$ is generated from a partition $P$ with three QI-groups $G_1 = \{$Ann, Bob$\}$, $G_2 = \{$Cate, Don$\}$, and $G_3 = \{$Ed, Fred, Gill$\}$. In addition, the adversary can infer that all sensitive values in $G_3$ must have not been modified by *Mask*. Otherwise, the distribution of sensitive values in $G_3$ must be adopted from another QI-group in Table II, which is impossible since neither $G_1$ nor $G_2$ has the same sensitive value distribution as $G_3$. On the other hand, the sensitive values in $G_1$ and $G_2$ may or may not have been modified by *Mask*. This leads to three different cases:

(1) *Both $G_1$ and $G_2$ have been modified.* This case is impossible; otherwise, the distributions of sensitive values in $G_1$ and $G_2$ should have been transformed to the same as in $G_3$, which is the only QI-group in $P$ that satisfies 2-diversity.

(2) *Either $G_1$ or $G_2$ has been modified.* In this case, one of $G_1$ and $G_2$ should contain two *dyspepsia* before modification (since *dyspepsia* is the only value in $V$), while the other one should have one *flu* and one *dyspepsia*. This results in 4 possible microdata instances, 3 of which assign *dyspepsia* to Ann.

(3) *Neither $G_1$ nor $G_2$ has been modified.* This leads to 4 possible microdata instances, 2 of which associate Ann with *dyspepsia*.

In summary, from the adversary's perspective, there exist 8 possible microdata instances that can be generalized into $T_{10}^*$, among which 5 instances associate Ann with *dyspepsia*. Therefore, the adversary has 5/8 posterior belief in the event that "Ann has dyspepsia".

| Age | Zipcode | Group ID |
|-----|---------|----------|
| 21 | 10000 | 1 |
| 27 | 18000 | 1 |
| 32 | 35000 | 2 |
| 32 | 35000 | 2 |
| 54 | 60000 | 3 |
| 60 | 63000 | 3 |
| 60 | 63000 | 3 |
| 60 | 63000 | 3 |

| Group ID | Disease |
|----------|-----------|
| 1 | dyspepsia |
| 1 | flu |
| 2 | bronchitis |
| 2 | gastritis |
| 3 | diabetes |
| 3 | dyspepsia |
| 3 | flu |
| 3 | gastritis |

| Age | Zipcode | Disease |
|---|---|---|
| [21, 27] | [10k, 18k] | flu |
| [21, 27] | [10k, 18k] | dyspepsia |
| [32, 54] | [35k, 60k] | gastritis |
| [32, 54] | [35k, 60k] | bronchitis |
| [32, 54] | [35k, 60k] | flu |
| 60 | 63000 | dyspepsia |
| 60 | 63000 | diabetes |
| 60 | 63000 | gastritis |

| Name | Age | Zipcode | Disease |
|------|-----|---------|---------|
| Ann | 21 | 10000 | flu |
| Bob | 27 | 18000 | dyspepsia |
| Cate | 32 | 35000 | gastritis |
| Don | 32 | 35000 | bronchitis |
| Ed | 54 | 60000 | flu |
| Fred | 60 | 63000 | dyspepsia |
| Gill | 60 | 63000 | diabetes |
| Hera | 60 | 63000 | gastritis |

| Age | Zipcode | Disease |
|---|---|---|
| [21, 32] | [10k, 22k] | dyspepsia |
| [21, 32] | [10k, 22k] | gastritis |
| [27, 36] | [18k, 37k] | flu |
| [27, 36] | [18k, 37k] | gastritis |
| [54, 60] | [60k, 63k] | bronchitis |
| [54, 60] | [60k, 63k] | flu |
| 60 | 63000 | dyspepsia |
| 60 | 63000 | diabetes |